\documentclass[aoas,nameyear,rotating,dvips]{arximspdf}
\usepackage{dcolumn}
\usepackage{graphicx}

\doi{10.1214/09-AOAS265}
\volume{3}
\issue{4}
\pubyear{2009}
\firstpage{1335}
\lastpage{1369}

\makeatletter
\newcolumntype{d}[1]{D{.}{.}{#1}}
\newcolumntype{k}[1]{D{;}{}{#1}}
\newproclaim{example}{Example}
\newtheorem{theorem}{Theorem}

  \let\sv@tabnotetext\tabnotetext
  \let\sv@tabnotemark@fmt\tabnotemark@fmt
   \long\def\legend#1{{\let\tabnote@indent\leavevmode\sv@tabnotetext[]{}{#1}}}

\makeatother

\begin{document}
\begin{frontmatter}

\title{Discovering influential variables:
A~method~of~partitions\thanksref{T1}}
\runtitle{Discovering influential variables: A~method of partitions}

\begin{aug}
\author[A]{\fnms{Herman} \snm{Chernoff}\ead[label=e1]{chernoff@stat.harvard.edu}\corref{}},
\author[B]{\fnms{Shaw-Hwa} \snm{Lo}\ead[label=e2]{slo@stat.columbia.edu}} \and
\author[B]{\fnms{Tian} \snm{Zheng}\ead[label=e3]{tzheng@stat.columbia.edu}}
\thankstext{T1}{Supported in part by by NSF Grant DMS-07-14669 and NIH
Grant R01 GM070789.}
\runauthor{H. Chernoff, S.-H. Lo and T. Zheng}
\affiliation{Harvard University, Columbia University and Columbia University}
\address[A]{H. Chernoff\\Department of Statistics\\ Harvard
University\\ Science
Center\\ Cambridge, Massachusetts 02138\\ USA\\\printead{e1}} 
\address[B]{S.-H. Lo\\ T. Zheng\\Department of Statistics\\ Columbia
University\\
New York, New York 10027\\ USA\\\printead{e2}\\\phantom{E-mail: }\printead*{e3}}
\end{aug}

\received{\smonth{9} \syear{2008}}
\revised{\smonth{4} \syear{2009}}

%
\begin{abstract}
A~trend in all scientific disciplines, based on advances in technology,
is the increasing availability of high dimensional data in which are
buried important information. A~current urgent challenge to
statisticians is to develop effective methods of finding the useful
information from the vast amounts of messy and noisy data available,
most of which are noninformative. This paper presents a general
computer intensive approach, based on a method pioneered by Lo and
Zheng for detecting which, of many potential explanatory variables,
have an influence on a dependent variable $Y$. This approach is suited
to detect influential variables, where causal effects depend on the
confluence of values of several variables. It has the advantage of
avoiding a difficult direct analysis, involving possibly thousands of
variables, by dealing with many randomly selected small subsets from
which smaller subsets are selected, guided by a measure of influence
$I$. The~main objective is to discover the influential variables,
rather than to measure their effects. Once they are detected, the
problem of dealing with a much smaller group of influential variables
should be vulnerable to appropriate analysis. In a sense, we are
confining our attention to locating a few needles in a haystack.

\end{abstract}

%

\end{frontmatter}
%

\section{Introduction}

\citeauthor{lo2002} (\citeyear{lo2002,lo2004}) introduced the backward haplotype-transmission
association (BHTA) algorithm, an efficient computationally intensive
method of detecting important genes involved in complex disorders. This
method, using haplotype information on multiple markers for affected
subjects and their parents, was applied to Inflammatory Bowel Disease
data [\citet{lo2004}]. In that application, a total of $235$
case-parent trios (each family contains an affected child and his/her
parents) and $448$ markers (variables) are included in the analysis.
Because the proposed method efficiently draws information from both
joint and marginal effects, interesting and novel scientific results
were obtained, some of them intriguing.


In order to accommodate different types of genetic data (such as in
case-control designs, e.g.), the method has been modified
recently to other genetic approaches using multiple markers [\citet{ionita2005}, \citet{zheng2006}]. A~brief summary of these methods and their
results appears in Supplement Section~S1 [\citet{onlinesuppl}].

In brief outline, the methods consist of subjecting a small randomly
selected group of markers to analysis to see which, if any of these,
seem to be moderately associated with the disease. A~measure $I$
[defined later in equation (\ref{for1})] that evaluates the amount of influence of
this set of markers is used to quantify their associations with the
disease. A~stepwise elimination process reduces this set to a smaller
set of possibly influential variables which are retained. By repeating
this process many times on randomly selected groups of markers, a
subset of markers is obtained which frequently appear to be associated
with the disease, and this subset is regarded as potentially involved
in the disorder.

An advantage of this method consists of avoiding a difficult direct
analysis involving hundreds or thousands of markers in favor of a
simple but effective analysis repeated many times. Another advantage is
that, as opposed to other methods depending mainly on marginal
information, this method can make use of both marginal and interactive
effects to yield more effective detections.

The~main idea applies much more generally than to special genetic
problems. In this paper a general version, which we shall call \textit{Partition Retention}, is proposed to deal with the problem of detecting
which, of many potentially influential discrete variables $ X_{s}$,
$1\leq s \leq S $, have an effect on a dependent variable $Y$ using a
sample of $n$ observations on $\mathbf{Z}=(\mathbf{X}, Y)$, where
$\mathbf{X} = (X_{1}, X_{2},\ldots, X_{S})$.

There exists a substantial literature, especially in engineering
journals, on feature and variable selection [\citet{Breiman2001},
\citet{Dash97}, \citet{Guyon03}, \citet{Koller96}, \citet{Ritchie01}], but much of it is directed
toward improving techniques in classification. A~set of variables that
are useful for classification purpose can be potentially very different
from the set of influential variables that we seek to identify in this
paper. In our text we will make comparisons with Random Forests [\citet
{Breiman2001}] and some comments on Multifactor Dimensionality
Reduction (MDR) [\citet{Ritchie01}]. In our discussion, we include some
comments on the interesting techinique of \citet{Koller96}.

In the background is the assumption that $Y$ may be slightly or
negligibly influenced by each of a few variables $X_{s}$, but may be
profoundly influenced by the confluence of appropriate values within
one or a few small groups of these variables.

At this stage the object is not to measure the overall effects of the
influential variables, but to discover them efficiently. Once these
variables have been detected, the problem of dealing with a much
smaller group of influential variables should be vulnerable to
appropriate analysis. In a sense we are confining our attention to
locating a few needles in a haystack.

The~object of this paper is to introduce the general approach, and to
indicate that there are many important variations of strategies which
may be worth exploring in order to increase the effectiveness for
finding influential variables and discarding \textit{impostors}.

Section \ref{sec2} provides a preliminary illustration of the approach with an
artificial example. This is followed by Section \ref{sec3} which gives a formal
presentation of terminology.


Sections \ref{sec4} and \ref{sec5} address the following major issues and the novel
advantages of our method with simple artificial examples in Section \ref{sec4}
and four more substantial ones in Section \ref{sec5}, where two are based on
real data:

\begin{enumerate}
\item As has been noted by \citet{Guyon03}, while one of a set of
influential variables may have no \textit{causal} effect by itself, it may
have an \textit{observable} marginal effect. That observable effect might
be small or negligible. Many current methods rely heavily on the
presence of strong observable marginal effects and are unlikely to
succeed if marginal observable effects are weak. Under certain
circumstances, some \textit{impostor} variables with no causal influence
may seem to have substantial marginal observable effects.

\item The~method we present is sensitive to the combined effects of
several influential variables when there are many potential influential
candidates. When the number of candidates is very large, our original
plan may not succeed in observing the combined effects of several
influential variables, and it may be necessary to \textit{thin} out the
set of candidates with a preliminary stage where all variables are
first considered one or two at a time. In a following stage our method
may \textit{resuscitate} influential variables that did not show up early.

\item Our method uses a measure of information related to the multiple
correlation (or $t$ test in the case of one explanatory variable). It
is more sensitive to influence than the correlation when applied to
several variables at a time.
\end{enumerate}

Section \ref{sec6} is a summary which also includes a discussion of the
comparison with Random Forests, and describes some aspects of an
interesting procedure by \citet{Koller96}. Finally, an appendix
contains some derivations and related results. Except for Appendices \hyperref[appendA]{A}
and \hyperref[appendB]{B}, the other parts (Supplement Sections~\mbox{S1--S3}) are included in the
online supplementary file [\citet{onlinesuppl}].

\section{Preliminary illustration}\label{sec2}

We introduce the partition retention (PR) approach and related
terminology and issues by considering a small artificial example.

\begin{example} \label{ex:1}
Suppose that an observed variable $Y$ is
normally distributed with mean $X_1X_2$ and variance 1, where $X_1$ and
$X_2$ are two of $S=6$ observed and potentially \textit{influential}
variables which can take on the values 0 and 1. Given the data on $Y$
and $\mathbf{X}=(X_1,\ldots,X_6)$, for $n=200$ subjects, the
statistician, who does not know this model, desires to infer which of
the six explanatory variables are causally related to $Y$. In our
computation the $X_i$ were selected independently to be 1 with
probabilities 0.7, 0.7, 0.5, 0.5, 0.5, 0.5.
\end{example}

The~approach is to partition the 200 observations into $2^6=64$ \textit{partition elements}, according to the values of ${\mathbf
X}=(X_1,X_2,\ldots,X_6)$ with $n_i$ observations in the $i$th element.
We introduce the measure
\[
I=n^{-1}\sum_i n_i^2(\bar{Y}_i-\bar{Y})^2,
\]
where $\bar{Y}=\sum_in_i\bar{Y}_i/n$ is the overall average of $Y$
and $\bar{Y}_i$ is the average of $Y$ in the $i$th element. We
consider $I$ to be a measure of influence based on how well the
partition separates the subjects into relatively homogeneous subsets.

To measure the influence of $X_1$ on $I$, we can repeat this process by
using the coarser partition depending on the other 5 variables, in
effect pretending that we do not have $X_1$ available. The~difference,
$D_1$, in the two values of $I$ is regarded as a measure of the
influence of $X_1$ on $Y$ in the presence of the other 5 variables. A~decrease in $I$ would suggest that $X_1$ has a substantial influence.
We could repeat this process for each of the other 5 variables. Our
procedure consists of discarding from consideration the variable for
which the $D$ value is least. We repeat this procedure with the
remaining 5 variables and continue discarding until we reach a step
where all the $D$ values are positive, at which time we \textit{retain}
the remaining variables.

We illustrate the method for a particular data set not presented here.
First we standardize $Y$ by subtracting the mean and dividing by the
standard deviation, a procedure we find convenient but not essential.
Then we obtain the value \mbox{$I=2.14$} when all 6 variables were considered.
Taking turns, eliminating one of these variables at a time gives us,
for the remaining five not eliminated, values of $I$ of 1.46, 1.57,
3.25, 3.32, 3.24 and 3.38, with corresponding $D$ values of 0.68, 0.57,
$-$1.11, $-$1.18, $-$1.10 and $-$1.24. Then we discard variable $X_6$ which led
to the smallest value of $D$, leaving us with a value of $I=3.38$.
Repeating this process on variables $X_1$ to $X_5$ leads to discarding
variable $X_4$ with $I=5.83$. An abbreviated history of this process is
presented in the first two rows of Table~\ref{tab:1} which give the
successive values of $I$ and the variables discarded at each stage.

The~next two rows of Table~\ref{tab:1} involve the same procedure
applied to the set of five variables $X_2, X_3, X_4, X_5, X_6$. The~following two rows treat the case where the variable $X_2$ is
originally omitted from the six. Finally, the last two rows treat the
case where only the last 4 noninfluential variables are considered in
the subset analyzed.

\begin{table}
\caption{History of the discarding procedure for four cases} \label{tab:1}
\begin{tabular*}{\textwidth}{@{\extracolsep{\fill}}ld{1.2}d{1.2}d{1.2}d{2.2}d{2.2}d{1.2}@{}}
\hline
\multicolumn{7}{@{}l}{Initial set: $\{1,2,3,4,5,6\}$} \\
$I$ before discarding &2.14 & 3.38 & 5.83 & 10.76 & 20.32 & 9.89 \\
Variable discarded & 6 & 4 & 3 & 5 & 1 & 2 \\ [6pt]
\multicolumn{7}{@{}l}{Initial set: $\{2,3,4,5,6\}$} \\
$I$ before discarding& 1.46 & 2.12 & 3.34 & 5.49 & 9.89 & \\
Variable discarded & 5 & 6 & 3 & 4 & 2 & \\[6pt]
\multicolumn{7}{@{}l}{Initial set: $\{1,3,4,5,6\}$} \\
$I$ before discarding& 1.57 & 2.29 & 3.36 & 5.49 & 8.70 & \\
Variable discarded & 6 & 3 & 4 & 5 & 1 & \\ [6pt]
\multicolumn{7}{@{}l}{Initial set: $\{3,4,5,6\}$} \\
$I$ before discarding& 1.00 & 1.12 & 1.13 & 1.01 & & \\
Variable discarded & 6 & 3 & 5 & 4 & & \\
\hline
\end{tabular*}
\end{table}

When the influential variables $X_1$ and $X_2$ are in the subset
subject to the process, they end up as the last items to be discarded.
When both are present the initial value of $I$ tends to be larger than
when only one is present, and when none are present the initial value
of $I$ is still smaller. In the first case our plan retains both
influential variables. In the next two cases $I$ increases as we
discard, and our retention strategy retains only the last variable
kept, 2 and 1 respectively. For the case where the discarding process
starts with only the ``unimportant'' $\{X_3, X_4, X_5, X_6\}$, $I$ has
the lowest initial value and does not grow much as variables are discarded.

The~strategy of retaining all variables when all $D$ values are
positive, that is, when $I$ starts to decrease, would lead to
retaining variables $X_4$ and $X_5$ in the fourth case. With the
relatively small initial value of $I=1.00$, it might be a good idea to
retain none of the variables being studied. In other words, our
strategy for retaining variables could be reconsidered. In fact, as we
shall note later, values of $I$ substantially greater than 1 signify
possible influence, and the values of $I$ at the stopping times were
20.32, 9.89, 8.70 and 1.13, in these four situations above. The~relatively modest value of $I$ at the stopping time in the fourth case
could be regarded as a signal to not retain the remaining variables.

Because we will be dealing with many candidate variables in more
realistic problems, our plan is to take small random subsets of the
variables under consideration and subject these to a reduction scheme
similar to the one described above. If the retention rate for
influential variables will be greater than for noninfluential
variables, the influential ones will show up more often in many
repetitions of this process, and will be discovered by their high
retention rates.

Although $X_1$ has no marginal \textit{causal influence} by itself, the
third case shows that it has a marginal \textit{observable effect} which
may also be detected by a simple $t$ test. As we shall see, applying
the $t$ test on each candidate variable is computationally cheap, and
may locate influential variables with a strong marginal observable
effect. But, in the case of many candidate variables, it will allow
some noninfluential variables to behave as \textit{impostors}. The~$t$
test may not be very efficient in detecting observable effects which
depend on interactions, and may fail to discriminate against some of
the impostors. However, for problems where $S$, the number of
potentially influential variables under consideration, is not very
large, we may be able to calculate the value of $I$ for all possible
pairs or even all possible triples, as a way of increasing the
sensitivity for detecting influential variables, for which the causal
effect depends largely on the interactions of groups of variables,
while discriminating against impostors.

\section{Formulation outline}\label{sec3}

If we select a subset or group of $m$ binary valued variables from
$\mathbf{X}= (X_{1}, X_{2},\ldots, X_{S})$, this subset defines a
partition $\Pi^{*}$ of the sample of $n$ observations into
$m_{1}=2^{m}$ subsets which we shall call partition elements, $\{A_{1},
A_{2},\ldots, A_{m_{1}}\}$, corresponding to the possible values of
the collection of these $m$ binary variables. For simplicity and
without causing confusion, we shall use $\{X_{1}, X_{2},\ldots,X_{m}\}
$ to denote the subset of selected variables. Each partition element
$A_{j}$ corresponds to a possibly empty subset of $n_{j}$ $Y$ values and
$\sum{n_{j}}=n$. Each nonempty partition element $A_{j}$ yields a mean
value $\bar{Y}_{j}$ and the overall mean\vspace*{1pt} is $\bar{Y}=\sum n_{j} \bar
{Y}_{j}/n $. Let
%
\begin{equation}\label{for1}
I_{\Pi^{*}}=n^{-1}\sum{n_{j}}^{2} (\bar{Y}_{j}- \bar{Y})^{2}.
\end{equation}

If $I_{\Pi^{*}}$ is unduly large, an expression to be explained later,
we suspect that some of the $m$ variables may have an influence on $Y$.

Suppose that we now introduce another binary variable from the original
set of $S$ potentially influential variables, which we shall call
$X_{0}$ for notational convenience. This leads to a more refined
partition $\Pi= \{ A_{jk} \dvtx  1\leq j\leq2^{m}, k=0,1\}$, where
$A_{j0}$ corresponds to that part of $A_{j}$ with $X_{0}=0$ and
$A_{j1}$ corresponds to that part of $A_{j}$ with $X_{0}=1$. Now let
$\bar{Y}_{jk}$ be the mean of the $n_{jk}$ elements in $A_{jk}$ and,
hence, $n_{j}=n_{j0}+n_{j1}$ and $n_j\bar{Y}_j=n_{j0}\bar
{Y}_{j0}+n_{j1}\bar{Y}_{j1}$. We
refer to $\Pi^{*}$ and $\Pi$ as the \textit{coarse} and \textit{refined}
partitions respectively. The~measure $I_{\Pi^{*}}$ is now replaced by
%
\begin{equation}\label{for2}
I_{\Pi}=n^{-1}\sum{n_{jk}^{2} (\bar{Y}_{jk}-\bar{Y})^{2}}
\end{equation}
and
%
\begin{equation}\label{for3}
D_{I}= \tfrac{1}{2}  ( I_{\Pi}- I_{\Pi^{*}})
\end{equation}
can be regarded as a measure of how much $X_{0}$ contributes in
influence on $Y$ in the presence of $\mathbf{X}= ( X_{1}, X_{2},\ldots,
X_{m} ).$ It is easy to see that
%
\begin{equation}\label{for4}
D_{I}= - n^{-1}\sum
n_{j0}n_{j1}(\bar{Y}_{j1}-\bar{Y})(\bar{Y}_{j0}-\bar{Y}).
\end{equation}
Thus, $D_{I}$ tends to be negative when both means in the refined
partition elements tend to be on the same side of $\bar{Y}$ as in the
coarse partition element from which the refined elements came. We would
expect that if the new variable contributes influence on $Y$, then
$D_{I}$ would tend to be positive.

In Appendix \hyperref[appendA]{A}, we calculate the expectation of $D_{I}$ conditional on
the partition sample sizes, in a more general framework described at
the end of this section. This expectation consists of the difference of
two positive quantities plus one which is relatively small and can be
estimated. Neglecting this small term, we see that if the new variable
has no influence on $Y$, other than random noise, the expectation of
$D_{I}$ will be nonpositive, and strictly negative if there are some
influential variables in the selected subset $\{X_1, \ldots, X_m\}$.
On the other hand, if the new variable $X_0$ contributes influence on
$Y$ and the old ones do not, then the expectation of $D_{I}$ will be positive.

Our policy is not that of adding new variables to our group of $m$
variables, but one of deleting variables from an initial group. Thus if
we start with $m+1$ variables, we consider the effect, that is, $
D_{I}$, of using the coarser partition obtained by eliminating one of
the $m+1$ variables. The~one with the smallest $D_{I}$ is then
eliminated, and we repeat this procedure on the remaining $m$
variables. We may continue eliminating until we are satisfied by some
criterion (e.g., when all the remaining $D_{I}$ are positive),
that most of the remaining variables are \textit{good} candidates for
being influential and should be retained.

The~set of $m+1$ variables will be selected at random from the original
set of $S$ variables. The~retention procedure is to be carried out many
times. We can observe which of the original variables is retained with
an unusually high frequency among those retained, and use these for
further analysis.

Because we expect to repeat this procedure many times, the approach is
computationally intensive. It may be possible sometimes to use the
initial value of $I_{\Pi}$ to decide whether a randomly selected group
of $m$ variables is worth pursuing with the elimination scheme, thereby
avoiding the calculations required for the successive eliminations.
Sometimes, it may be sensible to stop after the first step in the
elimination process and select the variables which lead to large values
of $D_I$.

While our discussion was confined to binary valued explanatory\break
variables, there is no such essential limitation. In fact, the
applications of \citet{zheng2006} used SNP genotypes, which assume
three possible values, as explanatory variables. We could easily
partition based on discrete valued explanatory variables. Then, if
$X_0$ assumes a finite set of values, say, 1 to $r$, the equation for
$D$ must be adjusted to give
\[
D_I=-n^{-1}\sum_i\sum_{j<k}n_{ij}n_{ik}(\bar{Y}_{ij}-\bar{Y})
(\bar{Y}_{ik}-\bar{Y}),
\]
where the partition element $A_{ij}$ is that subset of $A_{i}$ where
$X_0=j$ and has $n_{ij}$ elements averaging $\bar{Y}_{ij}$. If we
define $W_{ij}$ as the sum of all $Y$ variables in the partition
element $A_{ij}$ and $W$ as the sum of all the $Y$ values, then we may
write
\[
D_I=-n^{-1}\sum_i\sum_{j<k}\bigl(W_{ij}-(n_{ij}/n)W\bigr) \bigl(W_{ik}-(n_{ik}/n)W\bigr)
.
\]

In those cases where the explanatory variables are continuous, the
investigator could select cutoff points to separate the possible values
into a few discrete subgroups. This process might involve subjective
decisions. Such subjective decisions could also be applied to a more
complicated case where a pair of discrete or continuous variables may
be assigned to a small number of discrete values depending on the
expert opinions of the investigator.

In Appendix \hyperref[appendA]{A}, we deal with two models. In the first, the \textit{random
Y} model, we assume that the distribution of $Y$ depends on $\mathbf{X}$
which may be random or may be selected in advance as part of an
experimental design. In the second, the \textit{specified Y} model, the
values of $Y$ are selected in advance. For example, in case-control
experiments we select the number of cases and controls and examine the
related values of $\mathbf{X}$.

An alternative measure of influence, one more aligned with standard
analysis of variance calculations, is given by
%
\begin{equation}
J_{\Pi}=n^{-1}\sum n_{jk} (\bar{Y}_{jk}-\bar{Y})^{2}.
\end{equation}
The~use of $J$ to compare two sets of $m>1$ variables for influence is
the same as using the multiple correlation of $Y$ on these variables.
For $m=1$ using $J$, the squared correlation coefficient and the
absolute value of Student's $t$ will give almost the same comparisons
when $m=1$ and $r=2$. The~same could be said for the chi-square
statistic and $J$ when the variable $X_i$ assumes more than 2 values or
$m\ge2$.

\section{General comments}\label{sec4}

Our object is to locate influential variables. Whatever method we use
there is always the possibility that, among the ones we characterize as
influential, there will be some that are impostors. When $S$ is large,
it may be necessary to go through several stages of an elimination
process, eliminating many of the candidate variables from consideration
at each stage.

Our major method is most effective when the subgroup of variables
randomly selected has a reasonable probability of containing
more than one of the interacting influential variables. But for this to
be the case, the size of the randomly selected group of variables $m$
should be a substantial portion of $S$, the number of candidate
variables. When the sample size $n$ is not very large, $m$ has to be
modest for our approach to be effective, for otherwise, there will be
many partition elements that are empty or have only one subject. For
example, if $n$ is 200, we would like to have no more than 50 to 100
partition elements. In the case of binary valued $X$ values, that means
that we should consider subgroups of 6 or 7 variables at a time. But if
$S$ is 1000, it is rarely the case that a randomly selected subgroup
of 7 variables will contain more than one of a small number of
interacting influential variables. In that case, we have to thin out
the set of competitive variables before we can hope to have the
advantage anticipated when our subgroup frequently has more than one of
the interacting influential variables.

One way to thin out the candidates is to apply $I$ or the $t$ test to
one explanatory variable at a time, and to concentrate energy on those
which indicate strong marginal observable effects. If $S$ is not too
huge, we may even consider all possible pairs and concentrate on those
variables which appear in many \textit{high ranking} pairs.

In the examples of Section \ref{sec5} we will show how the partition retention
method applied after thinning can \textit{resuscitate} influential
variables with\break mediocre ranking by marginal considerations.

An issue of importance is the relative powers of using $m=1$, $2$ and $7$.
Another issue is whether a variable, which is an impostor using one of
these methods, is likely to fail by some of the others. If that is the
case, then we can hope to weed out impostors by combining the various
techniques. Finally, when our analysis points to some likely candidates
for being influential, we would like to have some way of deciding how
plausible our results are. One way would be to add randomly selected
additional variables which should have no relationship to the dependent
variable, and see how their presence affects the various statistics
used. This approach does not seem to be as reliable as simply permuting
the values of the observed dependent variable $Y$. This latter approach
does not upset the relationships among the $S$ explanatory variables,
whereas the first proposal would require assuming independence or some
arbitrarily selected correlations.

We have used the word impostor to suggest that some explanatory
variables which are not causally related to the dependent variable tend
to be easily confused with influential variables. The~following simple
artificial example provides some insight on the circumstances that can
lead to impostors.

\begin{example} \label{ex:2}
There are three independent explanatory
variables $X_1$, $X_2$, $X_3$ which take on the values 0 and 1. Let
$Y=X_1X_2$, and in a sample of $n$ cases, $n_{ijk}=np_{ijk}$ is the
number of cases where $X_1=i,X_2=j,X_3=k$. We use the subscript $d$ to
replace the conventional dot to indicate summation over the
corresponding index. For example, $p_{11d}=\sum_kp_{11k}=
p_{110}+p_{111}$. (In small print the symbol $d$ is easier to read than
a dot.) We also relate the $p$ values with the probabilities they
estimate and so we may write $p_{11d}=p(X_1=1,X_2=1)=p(X_1X_2=1)=p(Y=1)$.
\end{example}

Using the partitions based on $X_1$, we obtain
%
\begin{equation}\label{for6}
I_{X_1}=2n(p_{11d}p_{0dd})^2
\end{equation}
and using the noninfluential $X_3$, we have
%
\begin{equation}\label{for7}
I_{X_3}=2n\bigl(p_{11d}(p_{111}/p_{11d}-p_{dd1})\bigr)^2.
\end{equation}

The~ratio $I_{X_3}/I_{X_1}$ depends on the ratio of $p(X_3=1|X_1X_2=1)-
p(X_3=1)$ to $p(X_1=0)$. In Supplement Section S2 [\citet
{onlinesuppl}], we prove that, assuming independence of $X_3$ and
$X_1X_2$, the asymptotic distribution of the first term of this ratio
has mean 0 and variance
$n^{-1}P(X_3=1)P(X_3=0)P(X_1X_2=0)/P(X_1X_2=1)$. Thus, the probability
that the random noninfluential variable $X_3$ will act as an impostor
is small if $n$ is large. However, when $S$ is very large and $n$ is
modest, there may be several impostors.

Note that in most real problems the dependent variable is typically not
completely determined by the causal variables. There is usually some
random variation and the signal to noise ratio is of consequence. In
our example there was no noise, and the signal to noise ratio is
effectively infinite. Even so, it is possible to have impostors.

The~following is an artificial example where two explanatory variables
determine the dependent variable but neither one has a marginal
observable effect.

\begin{example} \label{ex:3} This is a variation of Example~\ref
{ex:2}, where the data consist of $(Y,X_1,X_2)$ for $n$ observations,
and $Y=X_1X_2+(1-X_1)(1-X_2)$. Then
%
\begin{equation}\label{for8}
I_{X_1}=2n(p_{11}p_{01}-p_{10}p_{00})^2.
\end{equation}
\end{example}

If $(X_1,X_2)$ takes on the values $(1,1)$, $(1,0)$, $(0,1)$ and $(0,0)$ with
probabilities $q_1, q_0, q_0, q_1$, then the expression
$p_{11}p_{01}-p_{10}p_{00}$ is asymptotically normal with mean 0 and
variance $2q_0q_1/n$. In effect, the marginal observable effects of
$X_1$ and of $X_2$ are negligible even though both variables are
influential. This lack of marginal observable effect depends on a
certain amount of symmetry in the causal mechanism and on the
distribution of explanatory variables.

It is useful to observe that if a group of noninfluential variables are
distributed independently of a dependent variable $Y$ which is
standardized to have sample variance 1, then $I$ will be distributed
roughly like a weighted sum of independent chi-squares with one degree
of freedom. Also, the distribution of $J$ conditional on $m'$, the
number of nonempty partition elements, will have the approximate
distribution of a chi-square with $m'$ degrees of freedom divided by $n$.
More precise statements and derivations are presented in Appendix \hyperref[appendB]{B}.
These results provide a clue as to when a group of variables are likely
to contain some influential ones.

Some experimental results, one listed in Supplement Section S3\break [\citet
{onlinesuppl}], suggest that both of these measures tend to have
roughly the same ability to detect influential variables when the
number of partition elements is small and of comparable sizes. However,
for a special alternative to independence, the ratio of sensitivities
of $I$ and $J$ depend heavily on $m'\sum(n_i/n)^2$, which attains a
minimum of 1 when all the partition elements have equal sample sizes,
and a maximum close to m when most of the observations are concentrated
in one partition element. The~advantage of $I$ over $J$ depends on the
variance of the frequencies $n_i$ (see Supplement Section S3 [\citet
{onlinesuppl}] for detail).

At this time, we hesitate to present a specific program to carry out
our aim of detecting influential variables. Each applied problem has
special needs which may call for variations on the procedures we described.

\section{Examples}\label{sec5}



In this section we present four examples. One is an extension of
Example~\ref{ex:3} of the last section and involves 10 influential
variables. Another is a more realistic one featuring two small groups
of influential variables. Two are based on a real data set for
Rheumatoid Arthritis. A~major advantage of the artificial ones is that
\textit{truth} is known and the properties of the methods can be evaluated
for those examples. By simulation we can see how our methods respond as
parameters of the model in the example vary. We have the opportunity to
compare the results with those of Random Forests (RF), a method
pioneered by \citet{Breiman2001}. For examples based on real data, we
have to rely on supplementary information to determine the reliability
of our conclusions.

Our simulations give rise to a great deal of data. For the sake of this
presentation, devoted to introducing the partition retention approach,
we will occasionally omit some useful information in an attempt to
avoid overwhelming the reader. In particular, we depend heavily on
ranking the influential variables among all the candidates, and rarely
present the measures used for the ranking. Thus, the reader will seldom
see those situations where there is a precipitous drop in the measure
as one goes from one variable to the one ranked next.

Our comparisons often will involve rankings of variables based on
$|t|$, $I_1$, $I_2$, $I_{2f}$, $I_7$ and RF. Here $t$ is the Student's
$t$ test statistic and behaves very much like $I_1$, which is the
marginal measure $I$ based on $m=1$. For $I_2$, we rank the $S(S-1)/2$
pairs of variables using $I$ based on $m=2$. There is no unique way to
rank the importance or influence of the individual variables given this
ranking of pairs. Two alternatives suggest themselves. The~first,
somewhat ambiguously labeled $I_2$, is measured by the number of
variables that have appeared at least once in the ranking of the pairs
before the candidate appears. The~potential trouble with this method is
the possibility that a very strong candidate in one group of
influential variables will carry some noninfluential variables with it
before we see indications from influential variables in another group.
An alternative ranking, $I_{2f}$, depends on the number of times a
candidate variable appears in the $n_r$ most highly ranked pairs where
$n_r$ is a substantial portion of the number of pairs.

Given a data set, $|t|$, $I_1$, $I_2$ and $I_{2f}$ are determined. The~partition retention method with $m=7$, yielding $I_7$, is random since
it depends on the random sample of $n_s$ subsets of $m$ variables. Good
choices of $n_s$ would depend on how far apart are the frequencies of
retention of influential and noninfluential variables. While real
problems could use sequential methods to help select $n_s$, we have
generally settled almost arbitrarily on $n_s=30{,}000$ or $20{,}000$ for
many of the experiments presented here. Similarly, in comparisons with
random forests, we have taken the number of variables sampled at each
node, $m_t$, to be 7 and the number of trees $n_t$ to be $20{,}000$.
Results for random forests seem to be insensitive to variations in
these parameters. The~rankings for random forests is given by RF.

Example~\ref{ex:4} is an artificial example, representing an extension
of Example~\ref{ex:3} to deal with 10 well balanced influential
variables in a set of 500 variables with 400 observations on each. We
shall see that marginal methods give poor results.

\begin{example}\label{ex:4}
The~$S=500$ variables $X_s$ are binary
valued with values 0 and 1. The~first 10 are influential. The~number of
ones among these, $R$, is uniformly distributed from 1 to 9. The~subset
of $R$ of these 10 variables to be equal to one is chosen at random
with equal probability from among all such subsets. The~remaining 490
variables are independent and each is chosen to have its probability of
one to be uniformly distributed from 0.4 to 0.6. Given $R$, the
dependent variable, $Y$ is normally distributed with mean and variance
equal to $4(R(R-1)+(10-R)(9-R))$. Here, the sample size $n=400$.
\end{example}

The~ranks of the influential variables are listed in Table~\ref
{tab:2a} when the methods $|t|$, $I_1$, $I_2$, $I_7$ and RF are applied.

\begin{table}
\caption{Ranks of influential variables using $|t|$, $I_1$, $I_2$,
$I_7$ and RF. Notation ``$r$'' is short for rank} \label{tab:2a}
\begin{tabular*}{\textwidth}{@{\extracolsep{\fill}}lcccccccccc@{}}
\hline
\textbf{Vars} & \textbf{1}& \textbf{2}& \textbf{3}& \textbf{4}& \textbf{5}& \textbf{6}& \textbf{7}& \textbf{8}& \textbf{9}& \textbf{10} \\
\hline
$r|t|$ &162& 281& 363& 69& 370& 52& 493& 337& 183& 290 \\
$rI_1$ &159& 279& 361& 65& 369& 50& 493& 335& 183& 288 \\
$rI_2$ & \phantom{00}8& \phantom{00}9& \phantom{00}3& \phantom{0}1& \phantom{00}6& \phantom{0}5& \phantom{00}4& \phantom{00}7& \phantom{00}2& \phantom{0}18 \\
$rI_7$ & \phantom{0}39& 123& 144& 35& 130& 33& 154& 161& \phantom{0}67& \phantom{0}45 \\
$r$RF &127& 363& 213& 51& 208& 48& 221& 220& 186& 266 \\
\hline
\end{tabular*}
\end{table}

\begin{table}[b]
\caption{Ranks of noninfluential variables 11 to 20 by $I_1$ and
$I_2$} \label{tab:3}
\begin{tabular*}{\textwidth}{@{\extracolsep{\fill}}lcccccccccc@{}}
\hline
\textbf{Vars} & \textbf{11}& \textbf{12}& \textbf{13}& \textbf{14}& \textbf{15}& \textbf{16}& \textbf{17}& \textbf{18}& \textbf{19}& \textbf{20} \\
\hline
$rI_1$& 207& 290& 345& \phantom{0}88& 190& 374& 466& 251& 158& 321 \\
$rI_2$& 370& 247& 348& 258& \phantom{0}33& 152& \phantom{0}63& 386& \phantom{0}28& 343 \\
\hline
\end{tabular*}
\end{table}

In summary, $|t|$ and $I_1$ are in close agreement and the ranks they
give are essentially those of a random sample of 1 to 500. There are no
first order observable effects, which is to be expected given the
construction. On the other hand, $I_2$ is almost perfect in identifying
the influential variables. Only the tenth is superseded by 8 impostors.
The~$I_7$ and RF methods did not do as well as $I_2$, but better than~$I_1$. Furthermore, in this example $I_7$ seems to do considerably
better than RF, suggesting that RF is more dependent than $I_7$ on
strong marginal observable effects. The~same calculations were done on
the data sets consisting of the first 100 and the first 200
observations. The~results for $I_2$ deteriorate slightly as the sample
size decreases, allowing 13 impostors. For $I_7$ and RF we still seem
to do better than chance, but not by very much.

One way of testing for influence is to introduce noisy variables and
see what effect these have on $I_1$. Another is to make comparisons
with the methods applied after $Y$ is randomly permuted a number of
times. Since we know truth in this case, these methods are not
required, but we demonstrate a couple of exercises. In Table~\ref
{tab:3} we apply $I_1$ and $I_2$ to the variables 11 to 500 and see how
the noisy variables 11 to 20 are ranked. In Table~\ref{tab:4} we apply
$I_1$ and $I_2$ to variables 1 to 500 and see how variables 1 to 10 are
ranked after subjecting $Y$ to a random permutation.

\begin{table}
\caption{Ranks of variables 1 to 10 under $I_1$ and $I_2$
after Y is randomly permuted} \label{tab:4}
\begin{tabular*}{\textwidth}{@{\extracolsep{\fill}}lcccccccccc@{}}
\hline
\textbf{Vars} & \textbf{1}& \textbf{2}& \textbf{3}& \textbf{4}& \textbf{5}& \textbf{6}& \textbf{7}& \textbf{8}& \textbf{9}& \textbf{10} \\
\hline
$rI_1$& 251& 374& 485& 283& 392& 338& 333& 430& 265& 465 \\
$rI_2$& 306& 433& 412& 241& 311& 293& \phantom{0}57& 217& 277& 340 \\
\hline
\end{tabular*}
\end{table}

The~comparison between the results in Tables \ref{tab:3} and \ref{tab:4} with those of
$I_2$ in Table~\ref{tab:2a} is striking. The~comparison with those of
$I_7$ and RF are less striking but apparent. In a real data problem, a
number of such randomized variations of the original data set can be
used to estimate the false discovery rate when this method is applied
to real data. Such an application appears later in Example~\ref{ex:resusRA}.

We will now introduce Example~\ref{ex:5} which has two small groups of
3 and 4 influential variables among 1000 candidate binary valued
variables. We consider two major aspects. First we examine the average
behavior of some of the methods, as sample size and signal strength
change. Here we find that average ranks are not very informative, since
one case with a large rank will hide the fact that most of the time the
rank is small. Thus, we report both the average and median ranks and
values of $I$. Second, on the assumption that a better understanding of
the intrinsic variabilities due to the underlying model and due to the
analysis would come from looking at a few examples in detail, we also
study five data sets generated by one of the models. Here we explore
the ability, by using $I_{2f}$ or a variation of $I_7$, to resuscitate
influential variables previously neglected.

\begin{example} \label{ex:5}
The~vector $\mathbf{X}$ has 1000 components which assume the values 0 and
1. The~first 7 consist of two sets of influential variables which
interact slightly. The~dependent variable $Y$ is normally distributed
with mean $\mu$ and standard deviation $\sigma,$ where
\[
\mu=\max(\mu_1,\mu_2) + 0.1(\mu_1+\mu_2)
\]
and
\[
\sigma=\max(\sigma_1,\sigma_2)
\]
with $\mu_1=\mu_0X_1X_2X_3$, $\mu_2=1.5\mu_0X_4X_5X_6X_7$,
$\sigma_1=1+X_1X_2X_3$ and $\sigma_2=1+2X_4X_5X_6X_7$. The~binary
valued explanatory variables are independent of each other and take on
the value of 1 with probabilities 0.4, 0.5, 0.6, 0.35, 0.45, 0.55 and 0.65 for
the seven influential variables. The~probabilities for the remaining
993 variables are randomly uniformly selected in the range of 0.4 to
0.6. In this example, there is a slight interaction between the rare
dual effects of the two groups of influential variables. After the data
set is observed, $Y$ is normalized to have sample mean 0 and variance 1.
\end{example}

First we will describe results based on 400 simulations of 4 \textit{conditions}. Then we will explore in depth 5 cases for one of these
conditions. The~four conditions involve sample sizes 200 and 400, and
the values 4 and 6 for $\mu_0$. For the four conditions we had
subsamples of $m=7$. We start with the results using the marginal
methods based on one or two variables.

Table~\ref{tab:5} presents, for each influential variable, the rank it
gets among the 1000 variables when each is subjected to the $t$ test
and when each is evaluated by $I_1$. This table is based on 400 data
sets corresponding to each of 4 conditions.

\begin{table}[b]
\caption{Ranks of the influential variables using $|t|$ and $I_1$
based on a single variable. Means and medians of the ranks using 400
data sets with $S=1000$. Four cases involve $n=200$ and 400 and $\mu
_0=4$ and 6}\label{tab:5}
\begin{tabular*}{\textwidth}{@{\extracolsep{\fill}}ld{2.2}d{2.2}d{2.2}d{2.2}d{2.2}d{3.2}d{3.2}ccc@{}}
\hline
\multicolumn{1}{@{}l}{\textbf{Variable}} & \multicolumn{1}{c}{\textbf{1}} & \multicolumn{1}{c}{\textbf{2}} & \multicolumn{1}{c}{\textbf{3}} & \multicolumn{1}{c}{\textbf{4}} & \multicolumn{1}{c}{\textbf{5}} & \multicolumn{1}{c}{\textbf{6}} & \multicolumn{1}{c}{\textbf{7}} & \multicolumn{1}{c}{$\bolds{n}$} &\multicolumn{1}{c}{$\bolds{\mu_0}$}&\multicolumn{1}{c@{}}{\textbf{Statistic}}\\
\hline
Mean &14.95&37.74&87.53&61.46&110.33&168.91&232.51& 200 & 4 & $|t|$ \\
Median & 2.00& 5.00&22.50& 8.00& 24.00& 62.00&134.50& & & \\[3pt]
Mean &15.57&37.00&90.25&67.06&110.11&168.63&245.70& & &$I_1$\\
Median & 2.00& 4.00&25.00& 9.50& 24.00& 62.50&154.00& & & \\ [6pt]
Mean & 5.61&18.95&55.76&43.21& 81.55&130.26&200.07& & 6 & $|t|$ \\
Median & 1.00& 3.00&13.00& 5.00& 17.00& 47.00& 98.00& & & \\ [3pt]
Mean & 5.88&18.37&57.92&47.88& 81.33&129.88&212.23& & &$I_1$\\
Median & 1.00& 3.00&14.00& 7.00& 17.00& 47.00&109.00& & & \\ [6pt]
Mean & 1.82& 4.99&13.55& 7.44& 23.54& 51.12&116.78& 400 & 4 & $|t|$ \\
Median & 1.00& 3.00& 4.00& 3.00& 6.00& 15.00& 37.50& & & \\ [3pt]
Mean & 1.84& 4.86&14.19& 8.60& 23.32& 50.92&125.17& & &$I_1$\\
Median & 1.00& 2.00& 4.50& 3.00& 6.00& 15.00& 46.00& & & \\ [6pt]
Mean & 1.64& 3.97& 8.40& 4.86& 12.93& 28.37& 89.84& & 6 & $|t|$ \\
Median & 1.00& 3.00& 4.00& 3.00& 5.00& 9.00& 29.00& & & \\[3pt]
Mean & 1.66& 3.83& 8.78& 5.52& 12.73& 28.15& 97.81& & &$I_1$\\
Median & 1.00& 2.00& 4.00& 3.00& 5.00& 9.00& 33.00& & & \\
 \hline
\end{tabular*}
\end{table}

Note that the first group of influential variables gets better average
results than the second group, and within each group the variables with
lower probabilities of~1 tend to do better. Given that the first group
is influential about one eighth of the time, while the second is
influential about half as often, it is natural to expect that elements
of the first group will be easier to detect in spite of the somewhat
weaker mean signal (mean 4 instead of 6 when $\mu_0=4$, and 6 instead
of 9 when $\mu_0=6$). If we think of the other variables in a group as
providing support to a designated variable being tested, the variables
which show up less frequently are being more strongly supported by the
more prominent members of its group. That is a heuristic explanation
for the second phenomenon.

\begin{table}[b]
\tabcolsep=0pt
\caption{Means and medians, based on 400 data sets, of values of $I_2$
and the ranks, $r$, for a few or the 21 pairs of influential variables}
\label{tab:6}
\begin{tabular*}{\textwidth}{@{\extracolsep{\fill}}ld{3.2}d{3.2}d{4.2}d{3.2}d{5.2}d{5.2}d{5.2}d{4.2}@{}}
\hline
\textbf{Pairs} & \multicolumn{1}{c}{$\bolds{(1,2)}$} & \multicolumn{1}{c}{$\bolds{(1,3)}$} & \multicolumn{1}{c}{$\bolds{(2,3)}$}
& \multicolumn{1}{c}{$\bolds{(1,4)}$} & \multicolumn{1}{c}{$\bolds{(4,5)}$} & \multicolumn{1}{c}{$\bolds{(5,7)}$}
& \multicolumn{1}{c}{$\bolds{(6,7)}$} & \multicolumn{1}{c@{}}{\textbf{Final}}\\
\hline
\multicolumn{9}{@{}l@{}}{$n=200,\mu_0=4$} \\
Mean $I_2$ & 8.34 & 7.74 & 6.25 & 6.43 & 4.40 & 3.18 & 2.68 & 4.85 \\
Med. $I_2$ & 8.06 & 7.32 & 6.05 & 6.31 & 4.11 & 2.88 & 2.48 & 4.63 \\
Mean $r$ & 758.45 & 921.87 & 4897.03 & 392.86 & 13178.51 & 42790.07 &53546.21 & 1034.57 \\
Med. $r$ & 3.00 & 6.00 & 50.00 & 46.50 & 1215.00 & 5697.50 & 11065.50 &514.00\\ [6pt]
\multicolumn{9}{@{}l@{}}{$n=200,\mu_0=6$} \\
Mean $I_2$ & 9.79 & 8.91 & 7.46 & 7.57 & 5.39 & 3.87 & 3.15 & 5.60 \\
Med. $I_2$ & 9.51 & 8.57 & 7.09 & 7.60 & 5.22 & 3.49 & 3.01 & 5.44 \\
Mean $r$ & 226.14 & 278.58 & 2074.82 & 170.89 &7688.84 & 23337.26 &37028.30 & 507.66 \\
Med. $r$ & 2.00 & 3.00 & 25.00 & 16.00 & 460.00 & 2923.00 & 5688.00 &274.00 \\ [6pt]
\multicolumn{9}{@{}l@{}}{$n=400,\mu_0=4$}\\
Mean $I_2$ & 15.55 & 14.23 & 12.20 & 12.20 & 8.58 & 6.26 & 5.22 & 8.68\\
Med. $I_2$ & 15.44 & 14.08 & 12.07 & 12.11 & 8.37 & 5.85 & 4.86 & 8.47\\
Mean $r$ & 16.84 & 48.54 & 212.85 & 22.03 &1197.38 & 44268.99 & 9317.81& 177.03 \\
Med. $r$ & 2.00 & 2.00 & 4.00 & 4.00 & 101.00 & 1207.50 & 2167.00 &70.50 \\ [6pt]
\multicolumn{9}{@{}l@{}}{$n=400,\mu_0=6$}\\
Mean $I_2$ & 18.77& 17.32 & 14.54 & 14.67 & 10.23 & 7.41 & 5.95 & 10.21\\
Med. $I_2$ & 18.51 & 17.06 & 14.30 & 14.57 & 10.11 & 7.22 & 5.59 & 9.93\\
Mean $r$ & 14.01 & 27.80 & 87.34 & 7.61 & 841.75 & 3315.74 & 5092.48 &141.41 \\
Med. $r$ & 1.00 & 2.00 & 4.00 & 4.00 & 67.50 & 1096.50 & 2124.00 &47.50\\
\hline
\end{tabular*}
\legend{Note: The~column ``Final'' refers to the rank at which all 7 influential variables
have appeared at least once.}
\end{table}

The~two methods, using the $t$ test and $I_1$, give comparable results.
This partially supports our claim that the use of $I$ is expected to be
preferred to $J$ when there are many unevenly occupied partition
elements, but not otherwise. It seems that the increase of $\mu_0$
from 4 to 6 has less effect than the doubling of the sample size,~$n$.
Even if the mean grows very large, there is a limited range of
improvement for a fixed sample size. In a sense, the experiment where
the dependent variable is a two-valued deterministic function of the
influential variables corresponds to a problem of our type where the
standard deviation of $Y$ given $X$ is zero or where the mean is
effectively infinite. But even there we can not eliminate impostors
with a finite sample size.

We present some of the results from calculating the value of $I_2$ for
pairs of variables in Table \ref{tab:6}. More precisely, we calculate $I_2$ for
all 499,500 pairs, and rank them in descending order. The~value of
$I_2$ and the rank $r$ for a given pair of influential variables, say,
4 and 7, are obtained. Finally, we determine the rank, by which all of
the influential variables have shown up, and the corresponding value of
$I_2$. We carry out these calculations, calculating means and medians
for 400 simulated data sets for each of the four conditions, $n=200$
and 400, and $\mu_0=4$ and 6. For simplicity and to save space, in
Table~\ref{tab:6} we show the results for only~7 of the 21 pairs of
influential variables. For comparison purposes, keep in mind that the
distribution of $I$ for a set of noninfluential variables is
approximately that of a weighted sum of independent chi-square
variables with one d.f., and hence has mean about 1 and variance about $
\sum2(n_i/n)^2$.


Table~\ref{tab:6} shows that sample size has a large effect on
uncovering influence, and signal to noise ratio has a relatively small
effect. For these conditions, many pairs have precedence over the ones
where all the influential variables have finally shown up. This table
indicating that we need a meaningful way of using pairs to rank single
variables suggested our use of $I_{2f}$ to supplement $I_2$. In
Table~\ref{tab:7} we apply $I_{2f}$ to 200 data sets for the condition
$n=400$ and $\mu_0=4$ with the two values of $n_r=2000$ and 4000.
The~results using $I_{2f}$ are substantially better than those using
$I_1$ and $I_2$, and a little better than for $I_7$. In other
applications with 1000 variables, we tend to use $n_r=5000$ more or
less arbitrarily. Presumably there may be a rational way of selecting
an appropriate value of $n_r$, but this question has not yet been examined.

\begin{table}
\caption{Means and medians of the ranks of influential variables, using
frequency of appearance in the $n_r$ most high ranked pairs for 200
data sets. In this case $n=400$ and $\mu_0=4$}
\label{tab:7}
\begin{tabular*}{\textwidth}{@{\extracolsep{\fill}}ld{2.2}d{2.2}d{2.2}d{2.2}@{}}
\hline
& \multicolumn{2}{c}{$\bolds{n_r=2000}$}& \multicolumn{2}{c@{}}{ $\bolds{n_r=4000}$}\\[-6pt]
& \multicolumn{2}{c}{\hrulefill}& \multicolumn{2}{c@{}}{\hrulefill}\\
\textbf{Variable} & \multicolumn{1}{c}{\textbf{Mean}} & \multicolumn{1}{c}{\textbf{Median}} & \multicolumn{1}{c}{\textbf{Mean}} & \multicolumn{1}{c@{}}{\textbf{Median}} \\
\hline
1 & 1.86 & 1.00 &1.64 & 1.00 \\
2 & 4.29 & 2.00 &6.32 & 2.00 \\
3 & 10.62 & 5.00 &11.20 & 4.00 \\
4 & 6.60 & 3.00 &10.91 & 4.00 \\
5 & 15.04 & 6.00 &27.89 & 6.00 \\
6 & 16.74 & 8.00 &39.51 &12.00 \\
7 & 24.84 &12.50 &51.52 &28.50 \\
\hline
\end{tabular*}
\end{table}

Next, the partition retention method described in the early sections of
this report was applied for the case where $n=400$ and $\mu_0=4$. This
was applied with 20,000 random subsets of 7, for each of 200 data sets.
For each data set the influential variables were ranked according to
how many times they were retained. The~means and medians of these ranks
are presented in Table~\ref{tab:8}.

\begin{table}
\caption{Means and medians, based on 200 data sets, of the ranks of
each of 7 influential variables, ranked according to the number of
retentions in 20,000 samples of 7 by the Partition Retention Scheme
when $n=400, \mu_0=4$ and $m=7$}
\label{tab:8}
\begin{tabular*}{\textwidth}{@{\extracolsep{\fill}}lccccccc@{}}
\hline
\textbf{Variable} & \textbf{1} & \textbf{2} & \textbf{3} & \textbf{4} & \textbf{5} & \textbf{6} & \textbf{7} \\
\hline
Mean & 19.42 & 39.94 & 86.02 & 17.73 & 6.81 & 8.60 & 22.92 \\
Median & \phantom{0}5.00 & \phantom{0}8.00 & 16.50 & \phantom{0}7.00 & 3.00 & 3.50 & \phantom{0}6.00 \\
\hline
\end{tabular*}
\end{table}

These results seem to be worse than those using the $t$ test for the
first 3 variables but better for the latter 4. The~reason for this is
not clear, but the difference suggests that, used in tandem, the two
approaches will have some effect in detecting and eliminating impostors.

A~key question concerns how to take advantage of the partial
information gained from the marginal observable effects.
One way is to reduce the number of plausible candidate variables, so
that our methods can apply higher order interactions to help detect
influential variables. Another way is to use likely candidates to
resuscitate influential variables that have not yet shown up well. To
investigate these possibilities, we will concentrate on a few data
sets. For each of these we will apply various techniques to see how
well these methods work.

First, Table~\ref{tab:9} provides a list of the 30 most favored
candidates by $I_1$, $I_2$, $I_{2f}$, $I_7$ and RF, using $m=1$, $m=2$,
$m=2$ with $n_r=5000$, $m=7$ with $n_s=20{,}000$ and $m_t=7$ with
$n_t=20{,}000$. Note that 30 was selected mainly to facilitate
presentation, and that in most comparable problems a larger number
would usually be more appropriate at this thinning stage.

\begin{sidewaystable}
\tabcolsep=0pt
\tablewidth=\textheight
\tablewidth=\textwidth
\caption{30 most highly ranked variables from five methods for each of 5 data sets} \label{tab:9}
\begin{tabular*}{\textwidth}{@{\extracolsep{\fill}}cd{3.0}d{3.0}d{3.0}d{3.0}d{3.0}d{3.0}d{3.0}d{3.0}d{3.0}d{3.0}d{3.0}d{3.0}d{3.0}d{3.0}d{3.0}d{3.0}d{3.0}d{3.0}d{3.0}d{3.0}d{3.0}d{3.0}d{3.0}d{3.0}@{}}
\hline
\multicolumn{5}{@{}c}{\textbf{Data set 1}}&\multicolumn{5}{c}{\textbf{Data set 2}}&\multicolumn{5}{c}{\textbf{Data set 3}}&\multicolumn{5}{c}{\textbf{Data set 4}}
 &\multicolumn{5}{c@{}}{\textbf{Data set 5}}\\[-6pt]
\multicolumn{5}{@{}l}{\hrulefill}&\multicolumn{5}{c}{\hrulefill}&\multicolumn{5}{c}{\hrulefill}&\multicolumn{5}{c}{\hrulefill} &\multicolumn{5}{c@{}}{\hrulefill}\\
$\bolds{I_1}$ & \multicolumn{1}{c}{$\bolds{I_2}$} & \multicolumn{1}{c}{$\bolds{I_{2f}}$} & \multicolumn{1}{c}{$\bolds{I_7}$}
 & \multicolumn{1}{c}{\textbf{RF}} & \multicolumn{1}{c}{$\bolds{I_1}$} & \multicolumn{1}{c}{$\bolds{I_2}$} & \multicolumn{1}{c}{$\bolds{I_{2f}}$} &
\multicolumn{1}{c}{$\bolds{I_7}$} & \multicolumn{1}{c}{\textbf{RF}} & \multicolumn{1}{c}{$\bolds{I_1}$} & \multicolumn{1}{c}{$\bolds{I_2}$}
 & \multicolumn{1}{c}{$\bolds{I_{2f}}$} & \multicolumn{1}{c}{$\bolds{I_7}$} & \multicolumn{1}{c}{\textbf{RF}} & \multicolumn{1}{c}{$\bolds{I_1}$}
  & \multicolumn{1}{c}{$\bolds{I_2}$} &\multicolumn{1}{c}{$\bolds{I_{2f}}$} & \multicolumn{1}{c}{$\bolds{I_7}$} & \multicolumn{1}{c}{\textbf{RF}}
   & \multicolumn{1}{c}{$\bolds{I_1}$} & \multicolumn{1}{c}{$\bolds{I_2}$} & \multicolumn{1}{c}{$\bolds{I_{2f}}$} & \multicolumn{1}{c}{$\bolds{I_7}$}
    & \multicolumn{1}{c}{\textbf{RF}} \\
\hline
\phantom{00}6 & 4 & 5 & 3 & 4 & 4 & 4 & 1 & 1 & 4 & 1 & 1 & 1 & 2 & 1 & 2 & 1 & 1 &2 & 4 & 1 & 1 & 1 & 4 & 1 \\
\phantom{00}5 & 5 & 6 & 4 & 5 & 1 & 7 & 4 & 4 & 1 & 2 & 2 & 2 & 5 & 4 & 4 & 2 & 2 &6 & 2 & 3 & 2 & 3 & 1 & 5 \\
\phantom{00}4 & 6 & 4 & 5 & 6 & 8 & 1 & 7 & 3 & 8 & 4 & 3 & 4 & 1 & 2 & 472 & 3 & 4& 676 & 1 & 5 & 3 & 5 & 5 & 4 \\
\phantom{00}3 & 7 & 3 & 870 & 1 & 7 & 5 & 8 & 5 & 3 & 5 & 4 & 5 & 4 & 5 & 1 & 218 &472 & 472 & 472 & 2 & 593 & 2 & 163 & 3 \\
\phantom{00}2 & 1 & 2 & 913 & 3 & 3 & 584 & 3 & 628 & 7 & 984 & 915 & 984 & 3 & 984& 3 & 4 & 3 & 1 & 5 & 4 & 888 & 4 & 999 & 2 \\
870 & 268 & 870 & 288 & 2 & 628 & 8 & 628 & 8 & 628 & 3 & 676 & 3 & 314& 314 & 677 & 472 & 677 & 462 & 677 & 593 & 5 & 593 & 3 & 593 \\
\phantom{00}1 & 182 & 1 & 809 & 106 & 2 & 6 & 2 & 469 & 2 & 874 & 7 & 314 & 759 &207 & 6 & 462 & 6 & 4 & 3 & 163 & 4 & 163 & 2 & 163 \\
106 & 673 & 346 & 2 & 623 & 674 & 3 & 542 & 794 & 5 & 314 & 97 & 874 &984 & 358 & 676 & 22 & 676 & 668 & 814 & 462 & 7 & 462 & 593 & 48 \\
660 & 3 & 106 & 6 & 870 & 469 & 2 & 674 & 2 & 690 & 759 & 609 & 358 &874 & 874 & 5 & 962 & 5 & 3 & 668 & 873 & 462 & 999 & 462 & 873 \\
789 & 106 & 660 & 106 & 660 & 690 & 614 & 690 & 542 & 6 & 358 & 790 &759 & 358 & 175 & 462 & 6 & 462 & 677 & 6 & 999 & 732 & 873 & 661 & 999\\
346 & 341 & 623 & 251 & 789 & 542 & 269 & 469 & 7 & 611 & 730 & 984 &730 & 376 & 702 & 668 & 853 & 668 & 5 & 676 & 143 & 999 & 143 & 240 &462 \\
623 & 251 & 789 & 623 & 403 & 233 & 628 & 6 & 674 & 63 & 427 & 42 & 929& 427 & 957 & 100 & 627 & 100 & 956 & 100 & 48 & 941 & 48 & 873 & 79 \\
288 & 687 & 288 & 403 & 520 & 6 & 961 & 233 & 611 & 661 & 251 & 571 &427 & 966 & 730 & 571 & 682 & 248 & 571 & 462 & 661 & 163 & 888 & 721 &798 \\
800 & 2 & 520 & 673 & 195 & 611 & 818 & 5 & 813 & 542 & 376 & 5 & 717 &657 & 759 & 713 & 362 & 956 & 713 & 178 & 888 & 306 & 661 & 893 & 143 \\
573 & 820 & 573 & 789 & 288 & 603 & 674 & 603 & 661 & 469 & 929 & 168 &207 & 730 & 3 & 956 & 916 & 571 & 276 & 742 & 610 & 38 & 233 & 233 &467 \\
520 & 243 & 800 & 962 & 346 & 5 & 562 & 152 & 233 & 794 & 717 & 909 &251 & 499 & 337 & 248 & 540 & 703 & 703 & 224 & 233 & 976 & 287 & 499 &610 \\
809 & 913 & 962 & 660 & 809 & 661 & 153 & 611 & 603 & 750 & 207 & 771 &657 & 929 & 391 & 178 & 204 & 713 & 937 & 713 & 79 & 136 & 721 & 287 &888 \\
962 & 454 & 809 & 7 & 573 & 262 & 469 & 63 & 614 & 233 & 175 & 592 &775 & 308 & 987 & 937 & 629 & 853 & 100 & 248 & 941 & 869 & 79 & 34 &941 \\
944 & 463 & 165 & 346 & 944 & 405 & 872 & 262 & 858 & 674 & 775 & 767 &376 & 853 & 929 & 742 & 874 & 742 & 814 & 853 & 499 & 679 & 610 & 48 &233 \\
913 & 563 & 403 & 944 & 800 & 63 & 690 & 794 & 690 & 603 & 957 & 522 &175 & 280 & 775 & 703 & 677 & 606 & 886 & 909 & 346 & 79 & 6 & 143 &661 \\
165 & 800 & 913 & 628 & 48 & 152 & 294 & 661 & 767 & 423 & 455 & 376 &455 & 455 & 975 & 924 & 590 & 924 & 853 & 924 & 929 & 143 & 346 & 346 &470 \\
403 & 288 & 944 & 1 & 407 & 813 & 418 & 750 & 6 & 143 & 660 & 864 & 957& 207 & 280 & 596 & 5 & 596 & 178 & 956 & 128 & 312 & 893 & 732 & 137 \\
182 & 870 & 182 & 800 & 905 & 794 & 913 & 143 & 262 & 262 & 657 & 856 &592 & 649 & 717 & 853 & 124 & 178 & 596 & 937 & 240 & 315 & 929 & 79 &929 \\
\hline
\end{tabular*}
\end{sidewaystable}

\renewcommand{\thetable}{\arabic{table}}
\setcounter{table}{8}
\begin{sidewaystable}
\tabcolsep=0pt
\tablewidth=\textheight
\tablewidth=\textwidth
\caption{Continued}
\begin{tabular*}{\textwidth}{@{\extracolsep{\fill}}cd{3.0}d{3.0}d{3.0}d{3.0}d{3.0}d{3.0}d{3.0}d{3.0}d{3.0}d{3.0}d{3.0}d{3.0}d{3.0}d{3.0}d{3.0}d{3.0}d{3.0}d{3.0}d{3.0}d{3.0}d{3.0}d{3.0}d{3.0}d{3.0}@{}}
\hline
\multicolumn{5}{@{}c}{\textbf{Data set 1}}&\multicolumn{5}{c}{\textbf{Data set 2}}&\multicolumn{5}{c}{\textbf{Data set 3}}&\multicolumn{5}{c}{\textbf{Data set 4}}
 &\multicolumn{5}{c@{}}{\textbf{Data set 5}}\\[-6pt]
\multicolumn{5}{@{}l}{\hrulefill}&\multicolumn{5}{c}{\hrulefill}&\multicolumn{5}{c}{\hrulefill}&\multicolumn{5}{c}{\hrulefill} &\multicolumn{5}{c@{}}{\hrulefill}\\
$\bolds{I_1}$ & \multicolumn{1}{c}{$\bolds{I_2}$} & \multicolumn{1}{c}{$\bolds{I_{2f}}$} & \multicolumn{1}{c}{$\bolds{I_7}$}
 & \multicolumn{1}{c}{\textbf{RF}} & \multicolumn{1}{c}{$\bolds{I_1}$} & \multicolumn{1}{c}{$\bolds{I_2}$} & \multicolumn{1}{c}{$\bolds{I_{2f}}$} &
\multicolumn{1}{c}{$\bolds{I_7}$} & \multicolumn{1}{c}{\textbf{RF}} & \multicolumn{1}{c}{$\bolds{I_1}$} & \multicolumn{1}{c}{$\bolds{I_2}$}
 & \multicolumn{1}{c}{$\bolds{I_{2f}}$} & \multicolumn{1}{c}{$\bolds{I_7}$} & \multicolumn{1}{c}{\textbf{RF}} & \multicolumn{1}{c}{$\bolds{I_1}$}
  & \multicolumn{1}{c}{$\bolds{I_2}$} &\multicolumn{1}{c}{$\bolds{I_{2f}}$} & \multicolumn{1}{c}{$\bolds{I_7}$} & \multicolumn{1}{c}{\textbf{RF}}
   & \multicolumn{1}{c}{$\bolds{I_1}$} & \multicolumn{1}{c}{$\bolds{I_2}$} & \multicolumn{1}{c}{$\bolds{I_{2f}}$} & \multicolumn{1}{c}{$\bolds{I_7}$}
    & \multicolumn{1}{c}{\textbf{RF}} \\
\hline
195 & 287 & 7 & 520 & 82 & 143 & 405 & 858 & 152 & 858 & 68 & 649 & 14& 957 & 251 & 551 & 571 & 737 & 737 & 276 & 287 & 605 & 941 & 919 & 308\\
251 & 369 & 195 & 799 & 962 & 353 & 143 & 351 & 760 & 119 & 280 & 726 &975 & 175 & 376 & 814 & 862 & 814 & 28 & 703 & 893 & 800 & 128 & 128 &476 \\
827 & 702 & 673 & 573 & 165 & 717 & 93 & 405 & 693 & 601 & 975 & 874 &34 & 661 & 524 & 662 & 741 & 937 & 51 & 422 & 470 & 721 & 470 & 758 &302 \\
140 & 799 & 140 & 287 & 913 & 750 & 750 & 460 & 119 & 152 & 337 & 150 &499 & 251 & 880 & 909 & 676 & 456 & 456 & 662 & 721 & 893 & 240 & 136 &499 \\
\phantom{00}7 & 140 & 827 & 165 & 893 & 858 & 323 & 717 & 750 & 405 & 928 & 401 &660 & 767 & 447 & 737 & 34 & 224 & 741 & 571 & 136 & 75 & 283 & 929 &55 \\
673 & 216 & 251 & 82 & 827 & 601 & 252 & 767 & 63 & 813 & 499 & 459 &308 & 775 & 502 & 224 & 889 & 227 & 256 & 92 & 283 & 817 & 312 & 983 &893 \\
628 & 573 & 144 & 195 & 976 & 802 & 687 & 693 & 351 & 534 & 702 & 581 &380 & 126 & 455 & 456 & 956 & 869 & 498 & 728 & 137 & 815 & 137 & 137 &283 \\
\hline
\end{tabular*}\vspace*{12pt}
%
\tabcolsep=0pt
\tablewidth=\textheight
\tablewidth=\textwidth
\caption{Five new candidates for each of 5 methods on five data sets derived
from the use of $I_7$ on the top 10 candidates} \label{tab:10}
\begin{tabular*}{\textwidth}{@{\extracolsep{\fill}}cd{3.0}d{3.0}d{3.0}d{3.0}d{3.0}d{3.0}d{3.0}d{3.0}d{3.0}d{3.0}d{3.0}d{3.0}d{3.0}d{3.0}d{3.0}d{3.0}d{3.0}d{3.0}d{3.0}d{3.0}d{3.0}d{3.0}d{3.0}d{3.0}@{}}
\hline
\multicolumn{5}{@{}c}{\textbf{Data set 1}}&\multicolumn{5}{c}{\textbf{Data set 2}}&\multicolumn{5}{c}{\textbf{Data set 3}}&\multicolumn{5}{c}{\textbf{Data set 4}}
 &\multicolumn{5}{c@{}}{\textbf{Data set 5}}\\[-6pt]
\multicolumn{5}{@{}l}{\hrulefill}&\multicolumn{5}{c}{\hrulefill}&\multicolumn{5}{c}{\hrulefill}&\multicolumn{5}{c}{\hrulefill} &\multicolumn{5}{c@{}}{\hrulefill}\\
$\bolds{I_1}$ & \multicolumn{1}{c}{$\bolds{I_2}$} & \multicolumn{1}{c}{$\bolds{I_{2f}}$} & \multicolumn{1}{c}{$\bolds{I_7}$}
 & \multicolumn{1}{c}{\textbf{RF}} & \multicolumn{1}{c}{$\bolds{I_1}$} & \multicolumn{1}{c}{$\bolds{I_2}$} & \multicolumn{1}{c}{$\bolds{I_{2f}}$} &
\multicolumn{1}{c}{$\bolds{I_7}$} & \multicolumn{1}{c}{\textbf{RF}} & \multicolumn{1}{c}{$\bolds{I_1}$} & \multicolumn{1}{c}{$\bolds{I_2}$}
 & \multicolumn{1}{c}{$\bolds{I_{2f}}$} & \multicolumn{1}{c}{$\bolds{I_7}$} & \multicolumn{1}{c}{\textbf{RF}} & \multicolumn{1}{c}{$\bolds{I_1}$}
  & \multicolumn{1}{c}{$\bolds{I_2}$} &\multicolumn{1}{c}{$\bolds{I_{2f}}$} & \multicolumn{1}{c}{$\bolds{I_7}$} & \multicolumn{1}{c}{\textbf{RF}}
   & \multicolumn{1}{c}{$\bolds{I_1}$} & \multicolumn{1}{c}{$\bolds{I_2}$} & \multicolumn{1}{c}{$\bolds{I_{2f}}$} & \multicolumn{1}{c}{$\bolds{I_7}$}
    & \multicolumn{1}{c}{\textbf{RF}} \\
\hline
\phantom{00}7 &2 &7 &7 & 7&5 &690&5 &7 &413&7 &5 &7 & 7& 7& 7&676&7&703&462&128&535&128&869 &128\\
881&520&944&1 &513&985&674&985&985&782&369&314&369&369&3&703&677&703&100& 7&299&177&558& 7 &299\\
332&623&623&660&789&915&628&915&674&56&154&874&154&326&759&294&956&100&853&676&7 &999&869&505 &505\\
520&972&789&520&543&405&639&603&590&690&308&984&918&126&154&100&853&853&294&853&869&505&299&299& 7\\
543&288&968&623&322&687&794&405&6 &628&326&812&126&154&369&853&5&294&834&956&38 &55 &7 &977 &869\\
\hline
\end{tabular*}
\end{sidewaystable}

One of the methods of using a reduced list consists of applying the
partition retention method to a sample of variables, 3 of which are
selected from a reduced list of the top 10 candidates and 4 of which
are selected from the variables not in the list. This method increases
the probability of getting two influential variables in the sample, one
of which may not yet be in the reduced list. We use the reduced list of
the 10 top variables. These will show up in $3/10$ of the samples, while
the ones not in the list will only appear in about 4 out of 1000
times. While 20,000 seemed reasonably large for applying $I_7$ using
$m=7$, our candidates for resuscitation will only show up 80 times, and
will be paired with a given variable of the reduced list only about 24
times. Of course, if the reduced list has two variables of a group,
then a member of the group not in the reduced list may pair up with one
of the two more often. However, 24 plus or minus about 5, is not a good
basis for discriminating between influential variables and impostors.
We have used 100,000 trials to get more opportunities for observing
interactions, although both the 20,000 in the first application and
100,000 here seem a bit modest.

The~10 members of the reduced list are sampled frequently and are bound
to be retained often. But some are retained so markedly less often than
others that they deserve to be eliminated. For this presentation, we
hesitate to do so in order to avoid unnecessary complications. Instead,
we will simply adjoin the five next most frequently retained ones to
the list of 10. These five appear in Table~\ref{tab:10}. Because they
were at a disadvantage in the first resuscitation step, we eliminate
that disadvantage in the next step by selecting 3 of the 15 in addition
to 4 of the remaining 985 variables for each sample of 7. After this
step there may be a rearrangement of the top 15, and a few new
contenders may appear after these 15.

In Table~\ref{tab:11} we list the rankings of the influential
variables for each of the five methods, $I_1$, $I_2$, $I_{2f}$, $I_7$
and RF, and for each of the five data sets. These rankings are followed
by ud1 and ud2 which give the rankings after each of the two
resuscitation steps. In these examples, the resuscitation methods
almost always seems to improve the rankings of influential variables
and often succeed in making prominent those influential variables that
were poorly regarded in the first approach. Note that it is virtually
impossible for variables with rank greater than 10 to achieve a rank
less than 11 in the first resuscitation. Also, because we did not
discard poor performers in that first step, it is unlikely that a
variable with rank greater than 15 will achieve a rank less than 16 in
the second step. Nevertheless, these resuscitations provide an
opportunity to reorder the candidates.

An alternative approach to discarding impostors and resuscitating
poorly ranked influential variables is to take a relatively large
number of prospects and submit those to the $I_{2f}$ approach. This
alternative approach can not resuscitate a variable which fails to
appear in the list of prospects, and so it would pay to use a
relatively large number of such prospects for this reduction stage,
which can then be repeated with a smaller list. This method, when
applied to the modest list of 30 prospects, provided considerable
improvement on the relative rankings of the influential variables in
the list, but failed to resuscitate variables not on the list of 30 for
the data sets 3, 4 and 5. The~results appear in Table~\ref{tab:12}.

\begin{sidewaystable}
\tabcolsep=0pt
\tablewidth=\textheight
\tablewidth=\textwidth
\caption{Ranks of influential variables for five methods applied to 5
data sets. Initial ranks, $rI$and $r$RF, and ranks ud1 and ud2 after
resuscitations} \label{tab:11}
\begin{tabular*}{\textwidth}{@{\extracolsep{4in minus 4in}}ld{2.0}d{2.0}ccccd{2.0}ccccd{2.0}d{2.0}d{2.0}ccd{2.0}cd{2.0}d{3.0}d{3.0}ccccd{2.0}d{2.0}d{3.0}ccccccd{4.0}@{}}
\hline
&\multicolumn{7}{c}{\textbf{Data set 1}} & \multicolumn{7}{c}{\textbf{Data set 2}} &\multicolumn{7}{c}{\textbf{Data set 3}} & \multicolumn{7}{c}{\textbf{Data set 4}} &
\multicolumn{7}{c@{}}{\textbf{Data set 5}} \\[-6pt]
&\multicolumn{7}{c}{\hrulefill} & \multicolumn{7}{c}{\hrulefill} &\multicolumn{7}{c}{\hrulefill} & \multicolumn{7}{c}{\hrulefill} &\multicolumn{7}{c@{}}{\hrulefill} \\
\multicolumn{1}{@{}l}{\textbf{Var}}&
\multicolumn{1}{c}{\textbf{1}}&\multicolumn{1}{c}{\textbf{2}}&\multicolumn{1}{c}{\textbf{3}} &\multicolumn{1}{c}{\textbf{4}}&\multicolumn{1}{c}{\textbf{5}}&
\multicolumn{1}{c}{\textbf{6}}&\multicolumn{1}{c}{\textbf{7}}
&\multicolumn{1}{c}{\textbf{1}}&\multicolumn{1}{c}{\textbf{2}}&\multicolumn{1}{c}{\textbf{3}} &\multicolumn{1}{c}{\textbf{4}}&\multicolumn{1}{c}{\textbf{5}}&
\multicolumn{1}{c}{\textbf{6}}&\multicolumn{1}{c}{\textbf{7}} &
\multicolumn{1}{c}{\textbf{1}}&\multicolumn{1}{c}{\textbf{2}}&\multicolumn{1}{c}{\textbf{3}} &\multicolumn{1}{c}{\textbf{4}}&\multicolumn{1}{c}{\textbf{5}}&
\multicolumn{1}{c}{\textbf{6}}&\multicolumn{1}{c}{\textbf{7}} &
\multicolumn{1}{c}{\textbf{1}}&\multicolumn{1}{c}{\textbf{2}}&\multicolumn{1}{c}{\textbf{3}} &\multicolumn{1}{c}{\textbf{4}}&\multicolumn{1}{c}{\textbf{5}}&
\multicolumn{1}{c}{\textbf{6}}&\multicolumn{1}{c}{\textbf{7}}&
\multicolumn{1}{c}{\textbf{1}}&\multicolumn{1}{c}{\textbf{2}}&\multicolumn{1}{c}{\textbf{3}} &\multicolumn{1}{c}{\textbf{4}}&\multicolumn{1}{c}{\textbf{5}}&
\multicolumn{1}{c}{\textbf{6}}&\multicolumn{1}{c@{}}{\textbf{7}} \\
\hline
$rI_1$&7&5&4&3&2&1&28&2&7&5&1&16&13&4&1&2&6&3&4&389&50&4&1&5&2&9&7&191&1&4&2&5&3&37&351\\
ud1&4&6&5&3&1&2&11&2&6&3&1&11&20&4&2&1&5&3&4&199&11&2&1&4&3&10&6&11&2&3&1&5&4&48&13\\
ud2&6&5&4&3&2&1&9&2&6&5&1&8&16&3&2&1&5&3&4&85&7&2&1&5&3&10&6&15&1&3&2&5&4&25&15\\
[3pt]
$rI_2$&5&14&9&1&2&3&4&3&9&8&1&4&7&2&1&2&3&4&14&748&7&1&2&3&5&22&10&73&1&2&3&7&6&49&8\\
ud1&6&11&5&1&3&2&4&2&6&3&1&7&8&4&1&2&3&4&11&269&5&2&1&4&3&14&6&74&2&3&1&5&4&18&8\\
ud2&6&7&5&3&2&1&5&2&6&4&1&7&11&3&1&2&5&3&4&190&6&2&1&4&3&10&6&18&2&3&1&5&4&31&10\\
[3pt]
$rI_{2f}$&7&5&4&3&1&2&24&1&7&5&2&14&12&3&1&2&6&3&4&42&163&1&2&5&3&9&7&44&1&4&2&5&3&20&199\\
ud1&5&6&4&2&3&1&11&2&6&3&1&11&55&5&2&1&5&3&4&113&11&1&2&4&3&10&6&11&1&3&2&5&4&32&15\\
ud2&6&5&4&3&2&1&7&2&6&5&1&8&16&4&1&2&5&3&4&148&8&2&1&5&3&10&6&15&1&3&2&5&4&23&15\\
[3pt]
$rI_7$&22&8&1&2&3&9&18&1&9&3&2&4&22&11&3&1&5&4&2&842&44&5&1&9&7&11&2&270&2&7&6&1&3&41&135\\
ud1&12&5&4&2&3&1&11&2&5&3&1&7&15&11&1&2&5&3&4&204&11&2&1&4&3&18&6&190&2&3&1&5&4&24&11\\
ud2&7&6&4&3&2&1&5&2&6&4&1&7&10&3&1&2&5&3&4&40&8&2&1&5&3&16&6&215&2&3&1&5&4&20&15\\
[3pt]
$r$RF&4&6&5&1&2&3&53&2&7&4&1&8&10&5&1&3&15&2&4&982&454&3&2&7&1&5&10&882&1&5&4&3&2&63&1000\\
ud1&2&1&4&3&8&6&12&2&6&3&1&7&8 &4&1&2&12&3&4&320&11 &2&1&4&3&8&6 &12&2&3&1&5&4&20&14\\
ud2&2&1&5&3&4&6&14&2&6&5&1&7&10&3&1&2&5 &3&4&59 &8 &2&1&5&3&4&6 &14&1&3&2&5&4&20&15 \\
\hline
\end{tabular*}
\end{sidewaystable}

\begin{sidewaystable}
\tabcolsep=0pt
\tablewidth=\textheight
\tablewidth=\textwidth
\caption{Ranks of influential variables for five methods applied to
five data sets. Initial ranks r$I_1$, r$I_2$, r$I_{2f}$, r$I_7$, $r$RF
before and after resuscitation with $I_{2f}$ based on the top 30 ranked
variables} \label{tab:12}
\begin{tabular*}{\textwidth}{@{\extracolsep{4in minus 4in}}ld{2.0}d{2.0}ccccd{2.0}ccccd{2.0}d{2.0}d{2.0}ccd{2.0}cd{2.0}d{3.0}d{3.0}ccccd{2.0}d{2.0}d{3.0}ccccccd{4.0}@{}}
\hline
&\multicolumn{7}{c}{\textbf{Data set 1}} & \multicolumn{7}{c}{\textbf{Data set 2}} &\multicolumn{7}{c}{\textbf{Data set 3}} & \multicolumn{7}{c}{\textbf{Data set 4}} &
\multicolumn{7}{c@{}}{\textbf{Data set 5}} \\[-6pt]
&\multicolumn{7}{c}{\hrulefill} & \multicolumn{7}{c}{\hrulefill} &\multicolumn{7}{c}{\hrulefill} & \multicolumn{7}{c}{\hrulefill} &\multicolumn{7}{c@{}}{\hrulefill} \\
\multicolumn{1}{@{}l}{\textbf{Var}}&
\multicolumn{1}{c}{\textbf{1}}&\multicolumn{1}{c}{\textbf{2}}&\multicolumn{1}{c}{\textbf{3}} &\multicolumn{1}{c}{\textbf{4}}&\multicolumn{1}{c}{\textbf{5}}&
\multicolumn{1}{c}{\textbf{6}}&\multicolumn{1}{c}{\textbf{7}}
&\multicolumn{1}{c}{\textbf{1}}&\multicolumn{1}{c}{\textbf{2}}&\multicolumn{1}{c}{\textbf{3}} &\multicolumn{1}{c}{\textbf{4}}&\multicolumn{1}{c}{\textbf{5}}&
\multicolumn{1}{c}{\textbf{6}}&\multicolumn{1}{c}{\textbf{7}} &
\multicolumn{1}{c}{\textbf{1}}&\multicolumn{1}{c}{\textbf{2}}&\multicolumn{1}{c}{\textbf{3}} &\multicolumn{1}{c}{\textbf{4}}&\multicolumn{1}{c}{\textbf{5}}&
\multicolumn{1}{c}{\textbf{6}}&\multicolumn{1}{c}{\textbf{7}} &
\multicolumn{1}{c}{\textbf{1}}&\multicolumn{1}{c}{\textbf{2}}&\multicolumn{1}{c}{\textbf{3}} &\multicolumn{1}{c}{\textbf{4}}&\multicolumn{1}{c}{\textbf{5}}&
\multicolumn{1}{c}{\textbf{6}}&\multicolumn{1}{c}{\textbf{7}}&
\multicolumn{1}{c}{\textbf{1}}&\multicolumn{1}{c}{\textbf{2}}&\multicolumn{1}{c}{\textbf{3}} &\multicolumn{1}{c}{\textbf{4}}&\multicolumn{1}{c}{\textbf{5}}&
\multicolumn{1}{c}{\textbf{6}}&\multicolumn{1}{c@{}}{\textbf{7}} \\
\hline
$rI_1$ &7 & 5 & 4 & 3 & 2 & 1 & 28 & 2 & 7 & 5 & 1 & 16 & 13 & 4 & 1 &2 & 6 & 3 & 4 & 389 & 50 & 4 & 1 & 5 & 2 & 9 & 7 & 191 & 1 & 4 & 2 & 5& 3 & 37 & 351 \\
ud &7 & 5 & 1 & 2 & 3 & 4 & 12 & 1 & 8 & 2 & 3 & 12 & 10 & 4 & 1 & 2 &6 & 3 & 4 & \multicolumn{1}{c}{$-$} & \multicolumn{1}{c}{$-$} & 1 & 2 & 7 & 3 & 10 & 5 & \multicolumn{1}{c}{$-$} & 1 & 2 & 3 & 4 &5 & \multicolumn{1}{c}{$-$} & \multicolumn{1}{c@{}}{$-$} \\
[3pt]
$rI_2$ &5 & 14 & 9 & 1 & 2 & 3 & 4 & 3 & 9 & 8 & 1 & 4 & 7 & 2 & 1 & 2& 3 & 4 & 14 & 748 & 7 & 1 & 2 & 3 & 5 & 22 & 10 & 73 & 1 & 2 & 3 & 7 &6 & 49 & 8 \\
ud &9 & 6 & 4 & 1 & 2 & 3 & 10 & 1 & 8 & 5 & 2 & 9 & 10 & 3 & 1 & 2 & 6& 3 & 4 & \multicolumn{1}{c}{$-$} & 17 & 1 & 2 & 5 & 3 & 9 & 7 & \multicolumn{1}{c}{$-$} & 1 & 2 & 3 & 5 & 4 &\multicolumn{1}{c}{$-$} & 21 \\
[3pt]
$rI_{2f}$ &7 & 5 & 4 & 3 & 1 & 2 & 24 & 1 & 7 & 5 & 2 & 14 & 12 & 3 & 1& 2 & 6 & 3 & 4 & 42 & 163 & 1 & 2 & 5 & 3 & 9 & 7 & 44 & 1 & 4 & 2 & 5& 3 & 20 & 199 \\
ud &7 & 5 & 1 & 2 & 3 & 4 & 12 & 1 & 7 & 5 & 2 & 12 & 8 & 3 & 1 & 2 & 6& 3 & 4 & \multicolumn{1}{c}{$-$} & \multicolumn{1}{c}{$-$} & 1 & 2 & 6 & 3 & 10 & 7 & \multicolumn{1}{c}{$-$} & 1 & 2 & 3 & 4 & 5& 17 & \multicolumn{1}{c@{}}{$-$} \\
[3pt]
$rI_7$ &22 & 8 & 1 & 2 & 3 & 9 & 18 & 1 & 9 & 3 & 2 & 4 & 22 & 11 & 3 &1 & 5 & 4 & 2 & 842 & 44 & 5 & 1 & 9 & 7 & 11 & 2 & 270 & 2 & 7 & 6 & 1& 3 & 41 & 135 \\
ud &7 & 5 & 1 & 2 & 3 & 4 & 12 & 1 & 8 & 2 & 3 & 14 & 9 & 4 & 1 & 2 & 5& 3 & 4 & \multicolumn{1}{c}{$-$} & \multicolumn{1}{c}{$-$} & 1 & 2 & 7 & 3 & 10 & 5 & \multicolumn{1}{c}{$-$} & 1 & 2 & 3 & 4 & 5& \multicolumn{1}{c}{$-$} & \multicolumn{1}{c@{}}{$-$} \\
[3pt]
$r$RF &4 & 6 & 5 & 1 & 2 & 3 & 53 & 2 & 7 & 4 & 1 & 8 & 10 & 5 & 1 & 3& 15 & 2 & 4 & 982 & 454 & 3 & 2 & 7 & 1 & 5 & 10 & 882 & 1 & 5 & 4 & 3& 2 & 63 & 1000 \\
ud &7 & 5 & 1 & 2 & 3 & 4 & \multicolumn{1}{c}{$-$} & 1 & 7 & 5 & 2 & 12 & 11 & 3 & 1 & 2 &7 & 3 & 4 & \multicolumn{1}{c}{$-$} & \multicolumn{1}{c}{$-$} & 1 & 2 & 6 & 3 & 10 & 7 & \multicolumn{1}{c}{$-$} & 1 & 2 & 3 & 5 &4 & \multicolumn{1}{c}{$-$} & \multicolumn{1}{c@{}}{$-$} \\
\hline
\end{tabular*}
\legend{Note: ``$-$'' is used to represent ranks not observed.}
\end{sidewaystable}

\subsection{Applications to Rheumatoid Arthritis}
\label{sec:RA}

This section describes an application of the methods of this paper to a
real data set on Rheumatoid Arthritis in two \textit{examples}. The~first
is a brief summary and expansion of work in which some of us
participated [\citet{ding2007}], and applies $I_2$ to thin the large
set of available SNPs, $I_8$ on the reduced set, and random
permutations of the dependent variable to estimate false discovery
rates. In the second we apply the ideas of resuscitation to obtain some
additional results.

\begin{example}\label{ex_2}
Rheumatoid Arthritis (RA, MIM 180300) is known as a common disorder
with complex genetic etiology. In \citet{ding2007} the Illumina genome
scan on RA, originally studied by \citet{amos2006}, was analyzed as
part of the Genetic Analysis Workshop 15 [\citet{gaw15}]. The~Ilumina
genome scan consists of 5407 Single Nucleotide Polymorphism (SNPs)
genotyped from 642 Caucasian families. For the analysis, 349 unaffected
individuals were selected as ``controls'' and 474 RA patients as
``cases.'' The~analysis was carried out in two stages. It should be
noted that in dealing with SNPs, we have explanatory variables which
can assume three possible values. Also, as pointed out in Supplement
Section S1 [\citet{onlinesuppl}], the BGTA method used in \citet
{ding2007} is equivalent to that of the partition retention method
using $I$.
\end{example}

Because there was a large number of three valued explanatory variables,
and $I_1$ seemed to be nonproductive, the first stage consisted of
using $I_2$ to select the 707~SNPs which appeared in the 1000 top
ranking pairs. Then the partition retention scheme $I_8$ was applied to
these 707 SNPs using 70,000 randomly selected subsets of 8 SNPs. Each
subgroup retained was assigned the value of $I$ at the stopping time.
This process yields a sample of 70,000 values of $I$. We plan to select
the elements of those retained subgroups for which the value of $I$ is
above a certain threshold. To determine that threshold, we applied 50
permutations to the case-control labels, repeating the process
described above for the original data each time. This yields 3,500,000
values of $I$. For each value of $I$, there are a number of selected
subsets from the original data that have a larger stopping value $I$,
say, $M_1$. At the same value of $I$, we also calculated, for each
permutation $b$, the proportion of the 70,000 permuted $I$ values that
are greater than the given value of $I$, say,~$p_0^{(b)}$. The~false
discovery rate (FDR) at this given value $I$ is then estimated as\looseness=-1
\[
\mbox{fdr}(I)=\frac{ \operatorname{median}(p_0^{(b)})}{M_1/70{,}000}
\]
[\citet{benjamini1995}, \citet{yekutieli1999}]. We used the value of $I$ when
the FDR estimate reaches 30\% as the selection threshold. Elements of
subsets with stopping $I$ values which exceed this threshold were
selected. These consisted of 50 SNPs which are located within 39
distinct genes. We shall call these SNPs \textit{qualified} since one can
not claim that they are truly related to RA and not impostors without
additional evidence from biological or other studies.

\begin{figure}

\includegraphics{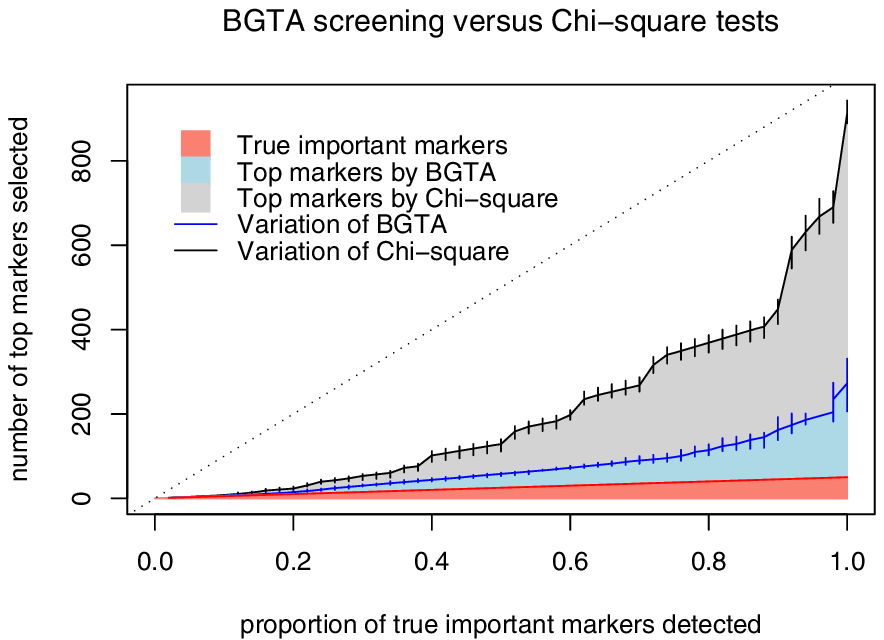}

\caption{Screening performance of BGTA and $\chi^2$ tests.
The~black and blue curves are average (out of five
simulated sets) number of top markers required to be selected in order
to attain a specific proportion of important markers. Vertical bars on
these curves indicate the maximum and minimum number of markers
required for the five simulations, which reflect the variability of the
retention method.}\label{fig:racurve}
\end{figure}

In this paper an additional procedure was carried out to determine how
well these qualified SNPs are ranked by $I_7$ in the presence of noise,
and how well this ranking compares with that of using the marginal
$\chi^2$-test. As illustrated in Supplement Section S1 [\citet
{onlinesuppl}], the $I_7$ approach is equivalent to the BGTA method
studied in \citet{zheng2006}. For this procedure the 50 qualified SNPs
are augmented by 950 additional SNPs selected at random from the
remaining 5357. For these additional SNPs the case-control designation
was permuted, while it was not for the 50. For example, if the
permutation moves the case row 16 to control row 35, our new row 35
will have the designation of case and the 50 SNPs will correspond to
those of the original row 16, while the remaining 950 will correspond
to those of row 35. In this way the structure of the values of the
unqualified SNPs is maintained while their relation with the dependent
variable is destroyed. This procedure was repeated 5 times with the
SNPs ranked by retention frequency using $I_7$ or BGTA (500,000
screenings for each data set) and by the $\chi^2$-test. In Figure~\ref
{fig:racurve} we plot the average proportion of SNP's ranked above a
given value against the average number of qualified SNPs ranked above
that value, for both methods. For example, by the time we retained 40
or 80\% of the qualified SNPs, we will have retained 100 by $I_7$
(BGTA) and 368 by the $\chi^2$-test. Each of these methods does
substantially better than pure chance in recognizing qualified SNPs. If
the qualified SNPs represented true effects, this figure would provide
an indication of an optimal cut off ranking, given the relative costs
of false discovery and of missing true relations using these methods.

\begin{example} \label{ex:resusRA}
In this example we apply resuscitation analysis on the real data on
Rheumatoid Arthritis (RA). We apply $I_{2f}$ with $n_r=50{,}000$ to
almost 15 million of the pairs of the 5407 SNPs. In other words, we
evaluated I for every pair, and ranked the individual SNPs according to
how often they showed up in the top 50,000 pairs. The~25 SNPs with the
top ranks were selected as the first reduced list.
\end{example}

To resuscitate influential SNPs not in this short list, the partition
retention method $I_7$ was applied to 2 million subsets with 3 members
from the 25 and 4 from the remaining SNPs. The~top ranking 50 SNPs from
this stage were then used for a second stage of $I_7$ with 2 million
subsets with 3 from the 50 and 4 from the remaining 5357. Table 13
displays the top 75 SNPs in this final resuscitated list,
38 of which are within 10 Mb of previously identified RA susceptibility loci.

\begin{table}
\caption{75 SNPs selected after two rounds of resuscitation.
Ranks from $I_{2f}$ and the two resuscitations~ud1~and ud2} \label{tab:13}
\begin{tabular*}{\textwidth}{@{\extracolsep{\fill}}lcd{2.0}d{3.1}d{3.1}l@{}}
\hline
&&&&& \textbf{Previously identified locus and} \\
\textbf{SNP}&\multicolumn{1}{c}{\textbf{Locus}}&\multicolumn{1}{c}{\textbf{ud2}}&\multicolumn{1}{c}{\textbf{ud1}}&
\multicolumn{1}{c}{$\bolds{rI_{2f}}$}&\textbf{reported significance (within 10 Mb)} \\
\hline
rs7534363 &1p36.3 &19&31&213&\citet{osorio}: $p=0.003$; \\
&&&&&\citet{cornelis}: $p=0.0035$; \\
&&&&&\citet{thompson}: $p=0.00585$;\\
rs2817594 &1p36.2 &33&32.5&236& ---\\
rs235256 &1p36.2 &28&32.5&151.5&---\\ [3pt]
rs569668 &1q42 &2&2&3&\citet{jawaheer}: $p=0.003$; \\
&&&&& \citet{osorio}: 0.04; \\
rs1389622 &1q44 &54&104&29& ---\\ [3pt]
rs300739 &2p25 &51&52&32&\citet{thompson}: note 1; \\
rs2685263 &2p25 &20&16&12& ---\\ [3pt]
rs6547142\tabnoteref[*]{tz} &2p12 &60&57&246&\citet{cornelis}: $p=0.041$; \\
rs1473357\tabnoteref[*]{tz} &2p12 &58&61&151.5& ---\\ [3pt]
rs921423 &2q11 &16&11&10& \\
rs7561232 &2q21 &10&27&205& \\
rs1402810 &2q21\mbox{-}22 &9&13&8& \\ [3pt]
rs970595 &2q33 &1&1&1&\citet{osorio}: $p=0.03$; \\
&&&&& \citet{cornelis}: $p=0.024$; \\
rs1921789 &2q33 &8&6&15& ---\\ [3pt]
rs3821280\tabnoteref[*]{tz} &2q37 &55&58.5&100&\citet{cornelis}: $p=0.0043$; \\ [3pt]
rs164466 &3p26 &11&9&9& \\
rs1385654 &3p12 &40&23&25& \\ [3pt]
rs4572747 &3q27 &63&55&273.5&\citet{cornelis}: $p=0.046$; \\
rs2067078 &3q27 &45&39&258& ---\\ [3pt]
rs881641 &4p16 &62&86&71&\citet{osorio}: $p=0.01$; \\ [3pt]
rs1424903 &5q11.2 &46&19&17& \\
rs1004531 &5q22\mbox{-}23 &3&5&4& \\ [3pt]
rs1560657\tabnoteref[*]{tz} &5q32\mbox{-}33 &52&162&27&\citet{cornelis}: $p=0.033$; \\ [3pt]
rs190129 &6p25 &25&45&125& \\ [3pt]
rs910516 &6p21 &31&14&13&\citet{osorio}: $p=6\mathrm{e}\mbox{-}5$; \\
&&&&& \citet{thompson}: $p=0.00127$; \\
&&&&& \citet{jawaheer}: $p=\mathrm{e}\mbox{-}12$; \\
&&&&& \citet{john}: $p=4\mathrm{e}\mbox{-}5$; \\
rs2277123 &6p21 &13&41&97& ---\\ [3pt]
rs508557 &6q13 &72&67&252&\citet{jawaheer}: $p=0.0028$; \\
rs6915493 &6q13\mbox{-}14 &21&29&213&\citet{jawaheer}: $p=0.0028$; \\
&&&&& \citet{john}: $p=0.006$; \\ [3pt]
rs2296412 &6q14\mbox{-}15 &49&34&291& \\
rs6934871 &6q15 &73&80&265& \\
\hline
\end{tabular*}
\end{table}

\renewcommand{\thetable}{\arabic{table}}
\setcounter{table}{12}
\begin{table}
\caption{Continued}
\begin{tabular*}{\textwidth}{@{\extracolsep{\fill}}lcd{2.0}d{3.1}d{3.1}l@{}}
\hline
&&&&& \textbf{Previously identified locus and} \\
\textbf{SNP}&\multicolumn{1}{c}{\textbf{Locus}}&\multicolumn{1}{c}{\textbf{ud2}}&\multicolumn{1}{c}{\textbf{ud1}}
&\multicolumn{1}{c}{$\bolds{rI_{2f}}$}&\textbf{reported significance (within 10 Mb)} \\
\hline
rs1873219 &6q15\mbox{-}16 &44&49&100&\citet{jawaheer}: $p = 0.01$; \\
rs4302647 &6q16 &14&28&241& ---\\
rs3827786 &6q16 &36&40&230.5& ---\\ [3pt]
rs2151913 &6q24\mbox{-}25 &12&8&11&\citet{cornelis}: $p=0.036$; \\ [3pt]
rs1852210 &7p13 &50&20&24& \\
rs691183 &7q32 &7&10&7& \\
rs1531381 &7q36 &67&70&79& \\ [3pt]
rs2442567 &8p23 &26&35&176.5&\citet{cornelis}: $p=0.040$; \\ [3pt]
rs766811 &8q24.2 &41&22&19& \\
rs751279 &9q22 &15&30&165& \\
rs715846 &9q22 &59&64&187& \\
rs2298033 &10p13 &56&51&176.5& \\ [3pt]
rs224136 &10q21 &71&68&95.5&\citet{jawaheer}: $p=0.0002$; \\ [3pt]
rs1649183 &10q26 &64&54&213& \\
rs4077638 &11q13 &42&38&181.5& \\
rs2276189 &11q24 &74&74&241& \\
rs6590098 &11q24 &53&56&192.5& \\ [3pt]
rs1558507 &12p13 &35&25&20&\citet{cornelis}: $p=0.0077$; \\ [3pt]
rs1517815 &12q21 &37&21&23&\citet{cornelis}: $p=0.0067$; \\ [3pt]
rs2070628 &12q24.3 &70&62&252& \\
rs866781 &12q24.3 &38&46&213& \\ [3pt]
rs4758930 &12q24.3 &39&36&291&\citet{john}: $p=0.05$; \\ [3pt]
rs1318725 &13q22 &17&17&22&\citet{cornelis}: $p=0.039$; \\
&&&&& \citet{john}: $p=0.03$; \\ [3pt]
rs3811310 &14q11.2 &29&43&187& \\
rs1570342 &14q11.2 &4&3&5& \\
rs1889387 &14q12\mbox{-}13 &18&12&14& \\
rs7149108 &14q12\mbox{-}13 &27&42&213& \\
rs4904723 &14q31\mbox{-}32 &48&47&236& \\
rs8005578 &14q32 &34&48&213& \\
rs7159412 &14q32 &75&79&213& \\
rs1365591 &15q12 &61&60&265& \\ [3pt]
rs1565863\tabnoteref[*]{tz} &15q14\mbox{-}15 &32&15&18&\citet{thompson}: $p=0.01634$; \\ [3pt]
rs3093291 &16p12\mbox{-}11.2 &5&4&2&\citet{cornelis}: $p=0.0080$; \\ [3pt]
rs7190151 &16q23 &69&58.5&143.5&\citet{cornelis}: $p=0.038$; \\
rs723919 &16q23 &22&18&16& --- \\ [3pt]
rs116719 &17q25 &57&66&221.5& \\
rs4479277 &17q25 &66&84&187& \\
rs6416862 &17q25 &24&44&173& \\
\hline
\end{tabular*}
\end{table}

\renewcommand{\thetable}{\arabic{table}}
\setcounter{table}{12}
\begin{table}
\caption{Continued}
\begin{tabular*}{\textwidth}{@{\extracolsep{\fill}}lcd{2.0}d{3.1}d{3.1}l@{}}
\hline
&&&&&\textbf{Previously identified locus and} \\
\textbf{SNP}&\multicolumn{1}{c}{\textbf{Locus}}&\multicolumn{1}{c}{\textbf{ud2}}&\multicolumn{1}{c}{\textbf{ud1}}
&\multicolumn{1}{c}{$\bolds{rI_{2f}}$}& \textbf{reported significance (within 10 Mb)} \\
\hline
rs879588 &18p11.2 &68&97&165&\citet{bache}: note 2; \\
&&&&& \citet{osorio}: $p=0.05$;\\ [3pt]
rs1661965 &19q13.3\mbox{-}.4 &23&26&192.5&\citet{kuroki}: $p=0.019$; note 3;\\
[3pt]
rs241605 &20p13 &43&24&21&\citet{osorio}: $p=1\mathrm{e}\mbox{-}4$; \\
&&&&& \citet{cornelis}: $p=0.030$; \\
rs761319 &20p12 &6&7&6& --- \\ [3pt]
rs1389157 &21q21 &47&50&265& \\
rs6517799 &21q21 &30&37&199.5& \\ [3pt]
rs5994180 &22q11.1\mbox{-}11.2 &65&53&273.5&\citet{bache}: note 2; \\
&&&&& \citet{osorio}: $p=4\mathrm{e}\mbox{-}2$; \\
&&&&& \citet{cornelis}: $p=0.019$; \\
&&&&&\citet{queiroz}: note 4.\\
\hline
\end{tabular*}
\legend{Notes: 1. Juvenile RA, LOD 6.0, stratified based on HLA-DRB1 presence.
2. The~mapping is based on chromosomal rearrangements in the Danish population on Juvenile RA.
3. Association to RA observed only for those that do not carry HLA-DRB1.
4. A~southern blot experiment revealed a gene IGLV8 being absent in RA
patients.}
\tabnotetext[*]{tz}{Identified SNP is $<$ 20 Mb from previously reported locus.}
\end{table}


In Table~\ref{tab:13} the SNPs are arranged according to the position
of the locus in the genome. Thirty eight of these 75 SNPs appear in 22
regions previously referenced in a publication.  The~remaining 37 SNPs have no
reference in the literature. Eleven of them end up among the top 25,
and 13 among the second 25. We find that the resuscitation has sent 9
from $I_{2f}$ rankings ranging from 97 to 241 to ud2 rankings in the
top 25.

To spare the reader from extensive tables, we have often presented
ranks and neglected frequencies and values of $I$. A~careful reading of
such extended tables would have made some things now obscure more
obvious. In particular, there is almost no chance that any of the first
25 SNPs would have a higher rank than 25 after the first resuscitation,
even if one shows up very poorly compared to the others in the first
25. The~second resuscitation gives a chance for those in the second 25
to push out some in the first 25. This happens for nine SNPs. There is
a possibility of using the results from nearby SNPs to give support to
a given locus, but we have not done so here.

In this paragraph we relate some of our results to the biological
literature. We use the term locus to represent a region of the genome
that has been identified as relevant to RA in the literature and may
contain several genes. Table 13 has 26 such loci, about 12 of which
were highly ranked by $I_{2f}$. These include lq44, 2p25, 2q33, 6p21,
6q24-25, 12p13, 12q21, 13q22, 15q14-15, 16p12-11.2, 16q23 and 20p13.
Some of these contain genes that are considered important in the
biological literature. For example, 6p21 contains HLA-DRB, which is
considered the most important RA gene identified to date. Also, 2q33
contains the important genes CTLA4, CD28 and STAT4, while 12pter-12p12
(centered at 12p13) contains CD4. The~locus lp36 harbors an important
RA susceptibility gene PADI4. At this locus we identified 3 SNPs which
required resuscitation to appear in the top 50 at ranks 19, 33 and 28.
The~gene LILR at 19q13.3-13.4 is known to be associated with RA
susceptibility among patients who are not HLA-DRB carriers [\citet
{kuroki}]. This may explain why this locus was discovered by
resuscitation and not in the initial $I_{2f}$ screening.

\section{Summary}\label{sec6}
We address a problem, expected in medical cases of complex diseases, of
a dependent variable influenced by one or a few small groups of
explanatory variables, when data is available on many such variables.
Our object is to detect these influential variables. Lo and Zheng
pioneered a method generalized here under the name Partition Retention.
This method samples $m$ of the $S$ variables many times and uses a
reduction process to retain a few of the $m$ variables. Those variables
that are retained most frequently are considered to be good candidates
for being influential. The~reduction process uses a statistic $I$ which
is considered to be a measure of information or influence for the set
of $m$ variables and $n$ is the sample size. On the null assumption
that the subset has no influential variables, the distribution of $I$
is approximately that of a weighted sum of independent chi-squares with
one degree of freedom.

When $S$, the number of variables, is large, the method is unlikely, in
its original form, to detect any variable that has a negligible
marginal observable effect. The~fact that an influential variable has
no marginal causal effect does not prevent it from having a marginal
observable effect. However, such effects can also be detected by other
first order methods. For example, a simple $t$ test will detect such an
effect, as will $I_1$ based on $m=1$. On the other hand, if $S$ is not
too huge, it is possible to consider second order interactions by
considering $I_2$ based on all pairs of variables. An alternative to
$I_2$ would be to evaluate the multiple correlations of the dependent
variable $Y$ on the pair of variables. There is some evidence that $J$
or the multiple correlation are as effective as the use of $I$ to rank
influence when $m$ is small. But when $m$ is large, we are likely to
have a large number of partition elements, many of which are empty or
have few members, and in that case the use of $I$ is more sensitive to
detect marginal observable effects.

The~rankings of pairs can be used to rank variables in several ways.
One is to to see how early influential variables are recognized when
$I_2$ is used to rank all pairs. An alternative which we prefer is to
rank all variables on how often they appear in the $n_r$ most highly
ranked pairs, where $n_r$ is a substantial fraction of approximately
$S^2/2$ pairs evaluated.

Assuming that $S$ is too large to consider all possible third order
interactions, we now have 4 methods. The~methods we label $I_1$, $I_2$,
$I_{2f}$ and $I_m$, based on one variable, two based on two variables,
and one based on $m$ variables, plus a few others similar to these but
using correlations. However, if $S$ is large, each of these methods may
pick out impostors among the plausible candidates for influential
variables. Part of our task is to discriminate against as many of the
impostors as possible. One approach may be to see how these various
techniques agree. The~assumption is that these methods provide tests to
determine influential variables, and insofar as the methods are
different, they will, in combination, provide a more difficult test for
an impostor than for a truly influential variable.

Another approach is that of using higher order methods on the
relatively few plausible candidates. We have used $I_{2f}$, with the
$S$ variables replaced by 30 candidates from each of the four
procedures on 5 distinct data sets with $S=1000$. We have also used a
variation of $I_7$ where 3 variables are selected from the top 10
candidates and 4 from the remaining $S-10$ variables. The~former method
does not make it possible to resuscitate influential variables not
among the 30 selected. The~latter method does make it possible.

For these second stage methods it would be feasible and sensible to
take a longer string of candidates to increase the probability of not
omitting influential variables. In fact, we used much longer strings of
plausible candidates in Example~\ref{ex:resusRA} on RA. It would also
be feasible to apply $I_3$ to a list of candidates, relatively small
compared with the original $S$. We have not done so here, nor have we
carried out another stage of reductions.

Our application of the partition retention system has been relatively
crude. The~desire to keep the presentation simple, without making
clever use of our knowledge of truth, led us to select numbers like 10
and 15 in our resuscitation scheme and other numbers almost
arbitrarily, without an attempt to show off the methods to advantage.
In the RA problem with real data, some necessary reasonable flexibility
in the choices was applied. It is worthwhile investigating various
strategies based on the use of $I$. It may save computing time if
subsets with initially small values of $I$ are ignored and not
subjected to the retention scheme. One alternative is to retain all
those variables which show a large positive value of $D_I$ on the first
step, and not bother with the rest of the reductions. Another is to
stop eliminating only if all $D_I$ values exceed a number depending on
the number of reductions that have taken place. In fact, one of the
weaknesses of the current method is that only one variable is retained
when $I$ increases with each stage of the reduction. This sometimes
permits a variable with a very strong signal to overwhelm other
influential variables that happen to be there.

Another valuable strategy adopted in our recent work [\citet{lo2008}],
which led to the discovery of interactions between various breast
cancer genes, was the use of the ratio of $I_2$ for a pair of genes to
the maximum of the values of $I_1$ for each of those genes compared to
a function of the maximum derived from the data.


The~Partition Retention (PR) method has some similarities with
Multifactor Dimensionality Reduction (MDR) [\citet{Ritchie01}] and
Random Forests (RF) [\citet{Breiman2001}]. MDR uses what we called
partition elements, but requires the dependent variable to be two
valued. It does such an intensive multifactor analysis on all possible
partitions that it is limited to problems with few explanatory
variables, about 20. It uses an error rate criterion, which we
conjecture might be improved by using the weighting implicit in $I$.

RF uses random subsets and is not limited to discrete explanatory
variables. Where PR is a backward recursion method which gets rid of
the worst candidates first, RF generates trees in a forward system,
that is vulnerable to confusion if the first choice is a poor one. In
other words, if the decision on the best first choice is not very good,
it is likely that future splits will not be useful. In PR, if the first
choice for deletion is not the least informative, the process is not
likely to be ruined.

In our Example \ref{ex:5}, we compared RF results with those of the other
techniques for the five special data sets. In those data sets which
exhibited strong first order observational effects, RF was comparable
to the other methods. Where the first order effects were not too
strong, RF seemed a little weaker. Resuscitation by $I_7$ and $I_{2f}$
worked well on RF, but not quite as well as for the other methods. As
far as we know, RF does not exploit the concept of resuscitating
variables that previously looked poor, but interact strongly with some
of those that looked good.

One of the referees brought the paper by \citet{Koller96} to our
attention. It has some interesting parallels to this manuscript. It
uses Kullback--Leibler information, for which $J$ is a first order
approximation, as a measure of influence, which we consider sensible,
but has two shortcomings. It requires that $Y$ be discrete, and it lacks
some of the advantages of $I$ over $J$. The~application of this information
is designed to attack the problem of causal variables without marginal
effect, by considering the effect of pairs. These pairs are employed in
an interesting way using so-called ``Markov blankets.'' However,
insofar as that method depends on those pairs, it is, like $I_2$ and
$I_{2f}$, deterministic, and does not allow for the resuscitation of
influential variables which require higher order interactions to be observed.

\begin{appendix}
\section{$\mathrm{E}(D_I)$}\label{appendA}

We derive expressions for the conditional expectation of $D_I$
given $\mathbf{n}$ for the random $Y$ model and the expectation
of $D_I$ for the specified $Y$ model.

\subsection*{Random-$Y$ model}

The~partition element $A_{ij}$ yields $n_{ij}$ independent
observations on $Y$, with mean $\mu_{ij}$ and variance $\sigma_{ij}^2$,
summing to $W_{ij}$. Let $\tilde{\mu}=n^{-1}\sum n_{ij}\mu_{ij}$
and $\tilde{\sigma}^2=n^{-1}\sum n_{ij}\sigma_{ij}^2$. We use the
tildes over
the Greek letters to remind ourselves that these depend on $\mathbf{n}$
and are not true parameters.

We calculate $\mathrm{E}(W_{ij}W_{ik})|\mathbf{n})=n_{ij}n_{ik}\mu_{ij}
\mu_{ik}$ for $j\neq k$,
$\mathrm{E}(W_{ij}W|\mathbf{n})=n_{ij}\sigma_{ij}^2+n_{ij}\mu
_{ij}n\tilde{\mu},$ and
$\mathrm{E}(W^2|\mathbf{n})=n\tilde{\sigma}^2+n^2\tilde{\mu}^2$.
Combining these expectations, we have
\[
\mathrm{E}(D_I|\mathbf{n})=-n^{-1}\sum_i\sum_{j<k}
n_{ij}n_{ik}[(\mu_{ij}-\tilde{\mu})
(\mu_{ik}-\tilde{\mu}) + n^{-1}(\tilde{\sigma}^2-\sigma
_{ij}^2-\sigma_{ik}^2)].
\]

The~term involving the variances is relatively small and can
be estimated. We will neglect it in this discussion. The~main
term can be rewritten. We replace the sum for $j<k$ by the
sum for $j\neq k$ and introduce $\mu_{ij}=\tilde{\mu}_i+\varepsilon_{ij}$
where $\tilde{\mu}_i=n_i^{-1}\sum n_{ij}\mu_{ij}$. Then
\begin{eqnarray*}
-2\mathrm{E}(D_I|\mathbf{n})&=&n^{-1}\sum_i\sum_{j\neq k}
[n_{ij}(\tilde{\mu
}_i-\tilde{\mu}+\varepsilon_{ij})
n_{ik}(\tilde{\mu}_i-\tilde{\mu}+\varepsilon_{ik})] \\
&=& H_1-H_2,
\end{eqnarray*}
where
$H_1= n^{-1}\sum_i [n_i(\tilde{\mu}_i-\tilde{\mu})
]^2$ and
$H_2=n^{-1}\sum_i\sum_j[n_{ij}(\tilde{\mu}_i-\tilde{\mu}
+\varepsilon_{ij})]^2$.

We may regard $\tilde{\mu}_i-\tilde{\mu}$ as representing the
effect of
$(X_1,X_2,\ldots,X_m)$ and $\varepsilon_{ij}$ as representing the
effect of $X_0$ in the presence of $(X_1,X_2,\ldots,X_m)$. Thus,
if the $m$ variables have no effect, $H_1$ will be zero, and $\mathrm{E}
(D_I|\mathbf{n})$
will be positive. If $X_0$ has no effect in the presence of the other
variables, $\varepsilon_{ij}$ will be zero and $\mathrm
{E}(D_I|\mathbf{n})$
reaches a minimal value which is nonpositive. The~greater the effect of
$X_0$ in the presence of $(X_1,X_2,\ldots,X_m)$, the more positive
$D_I$ tends to be. The~presence of influence in the $m$ variables tends
to diminish the effect of influence, if any, of~$X_0$.

\subsection*{Specified-$Y$ model}

Given that $Y$ assumes the values $y_1,y_2,\ldots,y_R$ with
frequencies given by $n^{(1)},n^{(2)},\ldots,n^{(R)}$, the
partition elements $ A_{ij}$ have $n_{ij}^{(r)}$ members equal to $y_r$,
where the $\mathbf{n}^{(r)}=\{n_{ij}^{(r)}\}$ are independent for
$1\leq r \leq R$ with
multinomial distributions $Mn(\mathbf{n}^{(r)},\mathbf{p}^{(r)})$,
and $\mathbf{p}^{(r)}=\{p_{ij}^{(r)}\}$.
The~number and probability for the partition elements $A_i$ are similarly
labeled $n_i^{(r)}$ and $p_i^{(r)}$. After normalization, $W=\sum n^{(r)}y_r=0$
and $I_{\Pi}=n^{-1}\sum W_{ij}^2$ and $I_{\Pi^*}=n^{-1}\sum W_i^2$.

We may write $W_i=\sum n_i^{(r)}y_r$ and $W_{ij}=\sum n_{ij}^{(r)}y_r$
and these
have expectations $\mathrm{E}(W_i)=\sum n^{(r)}p_i^{(r)}y_r$ and
$\mathrm{E}(W_{ij})=\sum n^{(r)}p_{ij}^{(r)}y_r$. To calculate
$\mathrm{E}(D_I)$, we need $\mathrm{E}(W_i^2)$
and $\mathrm{E}(W_{ij}^2)$, which involve the variances. We have
\[
\mathrm{E}(W_i^2)=(\mathrm{E}W_i)^2+\sum n_i^{(r)}p_i^{(r)}\bigl(1-p_i^{(r)}\bigr)y_r^2
\]
and
\[
\mathrm{E}(W_{ij}^2)=(\mathrm{E}W_{ij})^2+ \sum
n_{ij}^{(r)}p_{ij}^{(r)}\bigl(1-p_{ij}^{(r)}\bigr)y_r^2.
\]

Since the sums in the two expressions above are of order $n$ and
those of the squared expectations are of order $n^2$, we may
approximate $E(D_I)=\mathrm{E}(I_{\Pi})-\mathrm{E}(I_{\Pi^*})$ by
\[
\mathrm{E}(D_I)\approx n^{-1}\sum_i\biggl(\sum_j(\mathrm
{E}W_{ij})^2-(\mathrm{E}
W_i)^2\biggr).
\]

Let $p_{ij}=\sum n^{(r)}p_{ij}^{(r)}/\sum n^{(r)}$ and
$p_i=\sum n^{(r)}p_i^{(r)}/\sum n^{(r)}$. Let $e_{ij}^{(r)}=p_{ij}^{(r)}-p_{ij}$
and $e_i^{(r)}=p_i^{(r)}-p_{i}$. Then
\begin{eqnarray*}
\sum_i(\mathrm{E}W_i^2)&=&\sum_i\sum_{r,s}n^{(r)}y_rn^{(s)}y_s
\bigl(p_i^2+p_ie_i^{(s)}+
p_ie_i^{(r)}+e_i^{(r)}e_i^{(s)}\bigr) \\
&=&\sum_i\sum_{r,s}n^{(r)}y_rn^{(s)}y_se_i^{(r)}e_i^{(s)} \\
&=&\sum_i\biggl(\sum_r n^{(r)}y_re_i^{(r)}\biggr)^2.
\end{eqnarray*}
Also,
\begin{eqnarray*}
\sum_{ij}(\mathrm{E}W_{ij})^2&=&\sum_{i,j}\sum
_{r,s}n^{(r)}y_rn^{(s)}y_s
\bigl(p_{ij}^2+
p_{ij}e_{ij}^{(s)}+p_{ij}e_{ij}^{(r)}+e_{ij}^{(r)}e_{ij}^{(s)}\bigr)
\\
&=& \sum_{i,j}\sum_{r,s}n^{(r)}y_rn^{(s)}y_s e_{ij}^{(r)}e_{ij}^{(s)}
\\
&=& \sum_{i,j}\biggl(\sum_rn^{(r)}y_re_{ij}^{(r)}\biggr)^2.
\end{eqnarray*}

We have expressed $\mathrm{E}(D)$ as approximately the difference of
two positive
expressions, one of which involves $e_i^{(r)}$ which relates to the influence
of $X_1,X_2,\ldots,X_m$ on $Y$, and the other which involves $e_{ij}^{(r)}$
which relates to the combined influence of $X_0,X_1,\ldots,X_m$.

\section{Null distribution of $I$ and $J$}\label{appendB}

Consider the null distribution of $I$ for the random $Y$ model.
If the explanatory variables have no influence on $\mathbf{Y}$, we observe
$n$ independent identically distributed observations on $Y$
with $n_i=np_i$
allocated to partition element $A_i$. Then $I=n^{-1}\sum[n_i(\bar
{Y}_i-\bar{Y})]^2$
and $J=n^{-1}\sum n_i(\bar{Y}_i-\bar{Y})^2$. Suppose $Y$ has mean
$\mu$ and variance
$\sigma^2$. The~mean has no effect on the distribution of $I$ or $J$,
and $\sigma^2$ has only a multiplicative effect. Thus, there is no
loss of generality in assuming that $\mu=0$ and $\sigma=1$.

Naively assuming that $\bar{Y}=0$, we have the approximations that
$I=\sum p_i(n_i\bar{Y}_i^2)$ is distributed like
$\sum p_iV_i$, where
the $V_i$ are independent with approximately the chi-square
distribution with 1 d.f. as $n\rightarrow\infty$. A~similar
argument would have the distribution of $J$ approach that of
chi-square with $n'$ degrees of freedom
where $n'$ is the number of nonempty partition elements.

A~more precise derivation takes $\bar{Y}$ into account, but assumes
that all the $p_i$ are bounded away from 0 and 1 as $n$ gets large.
Let $Z_i=n^{-1/2}n_i(\bar{Y}_i-\bar{Y})$.
Conditioning on $\mathbf{n}=\{n_i\}$, the asymptotic distribution of
$\mathbf{Z}$ is $N(0,A)$, where $A=D(\mathbf{p})-\mathbf{p}\mathbf
{p}^T$ and $D(\mathbf{p})$
is the
diagonal matrix with elements $p_i$.

Since $I=\mathbf{Z}^T\mathbf{Z}$, the limiting distribution of $I$
is that of $\sum\lambda_i V_i$, where the $\lambda_i$ are
the eigenvalues of $A$. This singular matrix has one zero
eigenvalue, but the others are non-negative and add up
to the trace of $A$ which is $1-\sum p_i^2$. In most of
our ordinary applications $\sum p_i^2$ tends to be relatively
small and the naive approximation is a good fit. The~correction
for $J$ corresponds to the loss of one d.f.

In our applications we typically normalize $\mathbf{Y}$ so that it has
sample mean 0 and $n^{-1}\sum(Y_i^2)=1$. This normalization corresponds,
asymptotically, to replacing $\sigma$ by one.
On the other hand, our implicit assumption that all the $n_i$ are large
is really inappropriate for many of our applications where the
partition elements have a good number which are empty or singletons.
Nevertheless, it is easy to see that
$\mathrm{E}(I|\mathbf{n})=1-\sum p_i^2$, and it seems clear that a
more sophisticated
theorem will apply for these applications.

If we deal with the null distribution for the specified $Y$ model,
then the values of $Y$ in a given partition element with $n_i$
entries corresponds to a sample without replacement of $n_i$
observations from a finite population of $n$ elements with sum 0 and
sum of squares equal to $n$ after normalization. But then the sum
$W_i$ of the $Y$ values in partition element $i$ has $\mathrm
{E}(W_i)=0$ and
$\mathrm{E}(W_i^2)=n_i(1-(n_i-1)/(n-1)) \approx np_i(1-p_i)$.
Moreover, for $i\neq j$, the covariance $\mathrm{E}(W_iW_j)=-n_in_j/(n-1)$.
Since $I=\mathbf{W}^T\mathbf{W}/n$, the application of the Central Limit
theorem for sampling from finite populations repeats the analysis
for the random $Y$ model.
Once again, it is easy to see that $\mathrm{E}(I|\mathbf{n})=1-\sum p_i^2$.
We have proved the following:

\begin{theorem}
Conditioned on $\mathbf{n}$, the null distribution of $I$
when $Y$ is normalized is asymptotically that of a weighted sum of
independent chi-square variables, with non-negative coefficients
adding up to $1-\sum{p_i^2}$.
\end{theorem}

This applies to the null random $Y$ and the null specified $Y$ models,
under the standard conditions for the applicability of the Central
Limit theorem and the assumption that the elements of $\mathbf{p}$ are
bounded away from 0.

\end{appendix}

\section*{Acknowledgment}
 We would like to dedicate this to T. W. Anderson, a
pioneer in Multivariate Analysis, in honor of his 90th birthday. We
wish to thank the editors and referees for useful comments and references.

\begin{supplement}[id=suppA]
\sname{Supplement}
\stitle{Sections S1--S3}
\slink[doi]{10.1214/09-AOAS265SUPP}
\sdatatype{.zip}
\slink[url]{http://lib.stat.cmu.edu/aoas/265/Supplement.zip}
\sdescription{In the online supplements we detail several
previously-published methods as special cases of the
partition-retention approach (Section S1), the asymptotic distribution
of $p(X_3=1|X_1X_2=1)- p(X_3=1)$ discussed in Example 2 (Section S2)
and some discussion on relative efficiency of $I$ versus $J$ (Section S3).}
\end{supplement}

\printaddresses

\end{document}